\documentclass[utf8,utf8x,english]{frontiersSCNS} 

\usepackage[english]{babel}
\usepackage{url,hyperref,lineno,microtype}
\usepackage[onehalfspacing]{setspace}
\usepackage{multirow}
\usepackage{ulem}
\usepackage{subcaption}
\usepackage{caption}
\usepackage[utf8]{inputenc}

\def\keyFont{\fontsize{8}{11}\helveticabold} 
\def\firstAuthorLast{Bérard {et~al.}} 
\def\Authors{Rémi Bérard\,$^{1,2}$, Kremena Makasheva\,$^{2}$, Karine Demyk\,$^{1}$,
Aude Simon\,$^{4}$, Dianailys Nuñez Reyes\,$^{1}$, Fabrizio Mastrorocco\,$^{1}$, Hassan Sabbah\,$^{1,3}$, Christine Joblin\,$^{1,*}$}
\def\Address{$^{1}$IRAP, Université de Toulouse, CNRS, UPS, CNES, TOULOUSE, France\\
$^{2}$LAPLACE, Université de Toulouse, CNRS, UPS, INPT, TOULOUSE, France \\
$^{3}$LCAR-IRSAMC, Université de Toulouse, UPS, CNRS, TOULOUSE, France\\
$^{4}$LCPQ-IRSAMC, Université de Toulouse, UPS, CNRS, TOULOUSE, France}
\def\corrAuthor{Christine Joblin, IRAP,  9 av. du colonel roche, 31028, Toulouse Cedex 4, France}
\def\corrEmail{christine.joblin@irap.omp.eu}

\begin{document}
\onecolumn
\firstpage{1}

\title[Impact of metals on (star)dust chemistry]{Impact of metals on (star)dust chemistry: a laboratory astrophysics approach} 

\author[\firstAuthorLast ]{\Authors} 
\address{} 
\correspondance{} 

\extraAuth{}

\maketitle

\begin{abstract}

\section{}
Laboratory experiments are essential in exploring the mechanisms involved in stardust formation. One key question is how a metal is incorporated into dust for an environment rich in elements involved in stardust formation (C, H, O, Si). 
To address experimentally this question we have used a radiofrequency cold plasma reactor in which cyclic organosilicon dust formation is observed. Metallic (silver) atoms were injected in the plasma during the dust nucleation phase to study their incorporation in the dust.
The experiments show formation of silver nanoparticles ($\sim$15\,nm) under conditions in which organosilicon dust of size 200\,nm or less is grown. The presence of AgSiO bonds, revealed by infrared spectroscopy, suggests the presence of junctions between the metallic nanoparticles and the organosilicon dust. Even after annealing we could not conclude on the formation of silver silicates, emphasizing that most of silver is included in the metallic nanoparticles.
The molecular analysis performed by laser mass spectrometry exhibits a complex chemistry leading to a variety of molecules including large hydrocarbons and organometallic species. In order to gain insights into the involved chemical molecular pathways, the reactivity of silver atoms/ions with acetylene was studied in a laser vaporization source. Key organometallic species, Ag$_n$C$_2$H$_m$ (n=1-3; m=0-2), were identified and their structures and energetic data computed using density functional theory.
This allows us to propose that molecular Ag--C seeds promote the formation of Ag clusters but also catalyze hydrocarbon growth.

Throughout the article, we show how the developed methodology can be used to characterize the incorporation of metal atoms both in the molecular and dust phases. The presence of silver species in the plasma was motivated by objectives finding their application in other research fields than astrochemistry. Still, the reported methodology is a demonstration laying down the ground for future studies on metals of astrophysical interest such as iron.

\tiny
 \keyFont{ \section{Keywords:} laboratory astrophysics, stardust, dusty plasma,  organosilicon dust, silver nanoparticles, infrared spectroscopy, molecular analysis, density functional theory} 
\end{abstract}

\section{Introduction}

Interstellar space is populated by gas and dust made of (nano)grains of size less than typically 100\,nm.
A large fraction of this dust is expected to be formed in the cooling gas expelled from evolved stars and injected in the interstellar medium. The environments in which stardust forms include the envelopes of asymptotic giant branch (AGB) stars and the ejecta of supernovae.
They comprise a range of densities and temperatures and carry chemical complexity involving as main elements H, C, O and N but also less abundant species such as Si, S, and metals (Ti, Al, Mg, Fe).  
Infrared observations have led to identify two main classes of dust composition: carbon-based and oxide-based (silicates) dust, which was mainly attributed to the value of the C/O abundance ratio during their formation. Indeed, at thermal equilibrium, only the dominant of the two elements (C or O) is expected to remain available for dust formation, due to the consumption of the other element by the very stable CO molecule. For environments that are rich in oxygen, the formed solids would therefore be metal oxides (e.g. Al$_2$O$_3$) and silicates (e.g. Mg$_2$SiO$_4$ and MgSiO$_3$) and for environments rich in carbon one would expect the formation of silicon and titanium carbide as well as of solid carbon. But the analysis of observations suggests that other types of grains e.g., metallic iron grains, might be present \citep{Kemper2002}. This calls for a better understanding of the formation of stardust. However, this is a complex problem in which evolving chemical conditions are coupled with the changing physical conditions of the expanding shell or cooling ejecta.
 Most of the chemistry is therefore expected to be out of equilibrium and indeed molecular observations point to species whose observed abundances are not in agreement with equilibrium chemical models of stellar atmospheres (e.g detection of SiO and H$_2$O in C-rich environments; \citet{Schoier06}; \citet{Schoier11}; \citet{Cherchneff2012} and references therein). Considering these out-of-equilibrium conditions and the diversity of chemical elements that are involved, one could therefore expect a larger diversity of dust components, either as separate grains or combined in heterogeneous phases.

The conditions in which stardust forms differ from laboratory conditions. In particular, they involve much lower densities of species. In terms of gas-phase chemistry it implies that a reactive complex cannot be stabilized by 3-body reactions. Simulating the chemistry in circumstellar environments requires then to consider gas-phase chemistry and gas-grain interactions (once dust is formed) in low-density conditions and for a range of temperatures (typically from 2000\,K to $\sim$500\,K). Various techniques have been used to produce stardust analogues, which are based on laser ablation and laser pyrolysis \citep{Kroto85, Jaeger09}, flames \citep{Carpentier12},  pyrolysis \citep{Biennier09}, thermal evaporation \citep{Ishizuka2015} and dusty plasmas \citep{Contreras2013}. The conditions in these different experiments have been summarized in \cite{Martinez2020}. The latter article reports on the first experiments with a new setup, the Stardust machine, which presents the advantages of relatively well controlled conditions for the study of stardust formation. Still the machine is not designed to address the case of a too complex chemistry involving more than a couple of elements, which calls for preparatory experiments in order to tackle the impact of chemical complexity on dust formation and growth.

We report here a study focusing on the impact of the addition of a metal on dust formation in an environment in which the key elements involved in stardust formation (C, H, O, Si) are already present.  To address experimentally this question we have used a radiofrequency cold plasma reactor for which cyclic dust formation from the decomposition of an organosilicon precursor was observed  \citep{Despax12, Garofano2019}. In addition, the reactor has been designed to allow for the introduction of metallic (silver) atoms in the plasma during the dust nucleation phase. This is achieved by sputtering of a metallic target and has the advantage not to involve in dust formation additional fragments issued from the decomposition of metal-organic precursors. We have also performed complementary experiments using a newly developed laser vaporization source, in order to get further information on the gas-phase reactivity of silver atoms/ions and their small clusters with hydrocarbons. The dust composition was analyzed by infrared spectroscopy and the molecular composition by laser mass spectrometry. The experimental methodology was complemented by density functional theory calculations.
Throughout the article, we show as how the developed methodology can be used to characterize the incorporation of metal atoms into the molecular and dust phases. The presence of a silver electrode within the plasma reactor was motivated by objectives finding their application in other research fields in relation with the optical, electronic and biocide properties of silver nanoparticles. Still, we believe that the reported results provide general ideas about what could be expected for more astro-relevant metals such as titanium, aluminum, magnesium or iron. 

Our article is organized as follows. We first present our methodology followed by the results on infrared spectroscopy and molecular analysis. We include in the discussion the impact of Ag on the dust composition as well as its catalytic role on hydrocarbon molecular growth. We provide potential astrophysical implications. Finally, we conclude and present some perspectives.\\

\section{Material and Methods}

\subsection{Cold plasma experiments}

 Dust formation was achieved in a cold plasma successfully combining fragmentation of a precursor and sputtering of a metal target. The plasma reactor is a capacitively-coupled radio-frequency one (RF, 13.56\,MHz) with strong axial asymmetry that leads to a self-bias voltage (negative in this case) on the powered electrode and is at the origin of acceleration of the positive ions from the plasma towards this same electrode to bombard it and achieve sputtering. Typically, this type of plasma is used for deposition on a substrate of nanocomposite thin dielectric films with embedded metallic nanoparticles for different applications \citep{despax_deposition_2007, korner_formation_2010, milella_single_2014, Mak16}. In this work, we take advantage of this versatile cold plasma to focus on dust formation in the plasma gas-phase. The plasma was sustained at low gas pressure (5.3$~\times$~10$^{-2}$\,mbar) in argon gas (AirLiquid ALPHAGAZ 2 with a purity of 99.9995\%), injected continuously at 2.8\,sccm (standard cubic centimeters at standard pressure and temperature). The reactive gas used for dust formation was hexamethyldisiloxane (HMDSO, C$_6$H$_{18}$OSi$_{2}$, Sigma-Aldrich with a purity $>$ 99.5\%), injected by pulses. Stable cyclic dust formation in the plasma can exclusively be obtained when the precursor is injected by pulses \citep{Despax12, Garofano2019}. In the reported here experiments, the HMDSO gas pulses were injected during 1.5\,s each 5\,s, thus providing an averaged over the gas-pulses flow of 0.12\,sccm and leading to a total gas-mixture pressure of 5.6$~\times$~10$^{-2}$\,mbar. Such selection of the gas-pulse parameters allows simultaneous fragmentation of the precursor and sputtering of the metal target to provide the metal atoms. Silver (Inland Europe, with purity of 99.99\%) was used here, due to the multidisciplinary context of the study. A power of 30\,W was applied to the plasma to perform the experiments. Prior to each experiment, the reactor was mechanically cleaned with ethanol and pumped overnight. Further details on the experimental setup can be found in \cite{despax_deposition_2007} and \cite{Berard21}. The different stages of cyclic dust formation in the plasma gas-phase were monitored {\it in situ} by optical emission spectroscopy \citep{Despax12, Garofano2019}. For {\it ex situ} analysis such as electron microscopy, infrared spectroscopy and molecular analysis, as discussed below, the produced dust was collected on silicon substrates (provided by Sil'tronix).\\

\subsection{Laser vaporization experiments}
Gas-phase experiments were performed in the laser vaporization source (LVAP) of the PIRENEA 2 setup \citep{Bonnamy18}. LVAP is a laser aggregation source combining laser ablation with pulsed buffer gas in order to favor the formation of clusters from the vapor of ablated material. This technique is commonly used to produce gas-phase clusters \citep{Duncan12} and more precisely LVAP is based on the design described in \cite{Truong16}.  The second harmonic of a Nd:YAG pulsed laser (QUANTEL Q-smart 850) at 532~nm was focused on a silver rod (NEYCO 99.9994\% purity). At this wavelength, \cite{Duncan12} noticed that, in the case of silver, atomic species rather than clusters are formed. The typical energy of the laser pulse is 6--8~mJ with a repetition rate of 10~Hz. The rod is continuously rotated and translated by a motor to expose fresh surface to each laser spot. After each experiment, the silver rod is mechanically cleaned using a sponge and ethanol followed by 15~min in an ultrasonic bath filled with ethanol. Typical  background pressure in the source chamber is 3$~\times$~10$^{-7}$~mbar. Cooling of the laser plasma and clustering is produced by He gas with a backing pressure of 3.5~bar injected through a pulsed valve (Parker) with a repetition rate of 10~Hz and a pulse duration of 550~$\mu$s.
For reactivity studies, He gas containing 1\% of acetylene was used. In this configuration, the reactive gas and the Ag species created in the plasma interact near the ablation zone and in the growth channel of 1~cm long. The formed species are collected during 1-2~h of operation on a gold coated stainless steel disc (10~mm, TED PELLA PELCO AFM) located at 5~cm away from the ablation zone. The deposition disc is then extracted from the source for an {\it ex situ} analysis in the AROMA setup. The main operating parameters associated with the LVAP samples are given in Table~\ref{tab:samples}.\\ 

\subsection{Infrared spectroscopy}
The infrared (IR) spectra were recorded on the experimental setup ESPOIRS dedicated to the study of the optical properties of analogues of interstellar grains in the infrared and submillimeter range \citep[ESPOIRS;][]{Demyk17}. For these measurements, the ESPOIRS Fourier-transform spectrometer (Bruker VERTEX 70V) was equipped with a Globar source, a KBr beamsplitter and DLaTGS detector. The IR spectra were recorded under a controlled rarefied atmosphere of 0.5\,mbar with a 2\,cm$^{-1}$ resolution between 4000\,cm$^{-1}$ and 400\,cm$^{-1}$. A total of 256 to 1024 scans are co-added to ensure a good signal-to-noise ratio of the final spectra.  Plasma dust was collected on windows of intrinsic silicon substrates, because of their transparency to infrared light. The transmittance spectra are obtained after dividing the spectra of each sample by the spectrum of a blank silicon substrate recorded under the same conditions (pressure, temperature). The absorbance spectra are calculated from the transmittance spectra and the continuum is subtracted to facilitate comparison of the spectra. The evolution of the IR spectrum with temperature was followed by using the environmental cell of ESPOIRS (HT/HP Cell EN58J from Specac). To improve thermal exchange with the cell, we used a Si-intrinsic substrate for the high temperature measurements thicker than the one used for room temperature measurements. The two dust samples, for room temperature and for high temperature measurements, were collected during the same plasma experiment with the experimental conditions of sample P1 (see Table~\ref{tab:samples}). 
The sample was analyzed at room temperature, at 200$^{\circ}$C  and 500$^{\circ}$C. A temperature ramp of 5$^{\circ}$C/min was applied. At each selected temperature, the spectra were recorded every 30\,minutes over 2\,hours.\\

\subsection{Molecular analysis}\label{subsec:method_mole}
The molecular analysis has been performed on the Astrochemistry Research of Organics with Molecular Analyzer (AROMA) setup. It consists of a microprobe laser desorption ionization (LDI) source and a segmented linear quadrupole ion trap connected to an orthogonal time-of-flight mass spectrometer. In a first step, a pulsed (5 ns) near-infrared laser (Nd:YAG at 1064\,nm) is focused on the sample with a spot size of 300\,$\mu$m to cause rapid and localized heating. This favors thermal desorption over decomposition. The laser desorption fluence for this work varies in the 100-200\,mJ/cm$^{2}$ range. In a second step, a pulsed ultraviolet laser (fourth harmonic of an Nd:YAG at 266\,nm) intercepts perpendicularly the expanding plume of the desorbed molecules after 0.2 to 0.4\,$\mu$s and ionizes the organic molecules by (1+1) resonance-enhanced multi-photon ionization.  Due to its specific features, the ionization source is particularly adapted for the analysis of aromatic organic compounds. Aliphatic hydrocarbons are difficult to detect with AROMA since they involve larger ionization potentials that cannot be reached with our laser scheme \citep{Hanley2009}. The laser energy per pulse is optimized between 3 to 5\,mJ/pulse to have parent ions and not fragments dominating the mass spectra. Generated ions are driven to the orthogonal time-of-flight mass spectrometer trough an horizontal segmented ions trap. The detailed description of the setup can be found elsewhere \citep{Sabbah2017}.\\
\\

\subsection{Theoretical methods}\label{subsec:method_theo}
 In order to support the experimental data and provide first insights into C-H bond activation mechanisms, portions of potential energy surfaces (PES) for the complexes of interest were explored at the Density Functional Theory (DFT) level using the Gaussian16 suite of programs \citep{g16}.
 The wB97XD hybrid functional, which includes empirical dispersion and long-range corrections \citep{DFT_funct_2008}, was used in conjunction with the cc-pvtz basis set for C and H \citep{BS_CH_kendall_electron_1992}. The wB97XD functional was previously employed to model cycloaddition reactions on Ag and Ag$_2$ \citep{boz_ag-catalyzed_2016} in the context of modeling catalytic reactions on silver driving its biocide properties. The 4s,4p,4d and 5s electrons of Ag are described using a [6s5p3d] basis set \citep{martin_correlation_2001} associated to the Stuttgart relativistic effective core potential (ECP) \citep{Ag_BS_ECP_andrae1990a} that describes the 28 remaining core electrons. Basis sets and ECP were obtained from the Basis Set Exchange \citep{BSE1_feller1996a,BSE2_pritchard2019a,BSE3_schuchardt2007a}. We insist on the fact that the calculations presented in the present paper can be regarded as starting points for further theoretical studies aiming at rationalizing and complementing the experimental investigations. Only trends are searched for, and any systematic benchmark such as the one achieved for H$_2$ activation by Ag$_2$ and Ag$_3$ \citep{moncho_benchmark_2013} are out of the scope presently.
 
 Local DFT geometry optimizations were complemented by full harmonic frequency calculations  diagonalizing the Hessian matrix. Each stationary geometry was characterized as a minimum or a saddle point of first order by frequency calculations, which were also used to obtain the zero-point vibrational energies (ZPE). The search for the transition states (TS) was accomplished through the Synchronous Transit-Guided Quasi-Newton (STQN) search algorithm, followed by the Berny algorithm, and confirmed by the Intrinsic Reaction Coordinate (IRC) procedure, as implemented in Gaussian16.  Coupled-Cluster with Single and Double and perturbative Triple excitations (CCSD(T)) single point calculations on the DFT optimized geometries, using the same basis set, were occasionally performed to confirm the energetics. \\
 
\section{Results}

\begin{table}
\begin{tabular}{|c|c|c|c||c|c|c|}
\hline
\multicolumn{4}{|c||}{}&\multicolumn{3}{c|}{}\\
\multicolumn{4}{|c||}{Plasma samples}&\multicolumn{3}{c|}{LVAP samples}\\
\multicolumn{4}{|c||}{}&\multicolumn{3}{c|}{}\\
\hline
\multicolumn{4}{|c||}{}&\multicolumn{3}{c|}{}\\
\multicolumn{4}{|c||}{HMDSO (C$_6$H$_{18}$OSi$_{2}$) / silver}&\multicolumn{3}{c|}{ C$_2$H$_2$ / silver} \\

\multicolumn{4}{|c||}{}&\multicolumn{3}{c|}{}\\
\hline
&&&&&&\\

Name & Absorbed power & Dust & Collection time & Name & Laser energy & Collection time\\
&per unit area& generations & && & \\
&(W/m$^2$)&(number)&(min)&& (mJ/pulse)&(h)\\
 \hline
&&&&&&\\
P0* & 215 & 6 & 12 & & & \\
&&&&&&\\
P1  & 670 & 2 & 8 & LA1 & 8 & 1\\
&&&&&&\\
P2 & 643 & 5 & 23 & LA2 & 6 & 2 \\
&&&&&&\\
P3 & 590 & 3 & 13 & & & \\
&&&&&&\\
\hline
\end{tabular}
\caption{Short summary on the samples presented in this study. *Sample P0, also referred as reference sample, is made of organosilicon dust only, without silver component.}
\label{tab:samples}
\end{table} 

\subsection{Dust formation conditions and environment}\label{Sec:Dust}
The selected above plasma operating parameters (applied power, gas pressure, injected HMDSO amount and gas-pulse parameters) lead to the formation of successive dust generations in the plasma, approximately every 250\,s. Due to the dynamic character of the cyclic dust generation all plasma characteristics oscillate with the same frequency. The produced plasma has averaged values of its two main characteristics, the electron energy and the electron number density, of 2.2\,eV and 1.1$~\times$~10$^{10}$\,cm$^{-3}$, respectively \citep{Garofano2019, Berard21}. These values gradually vary during dust formation and dust escape from the plasma gas-phase. The electron energy and electron number density have opposite behavior. When dust grows in plasma the electron number density decreases (down to 1.0$~\times$~10$^{10}$\,cm$^{-3}$) because of the electron loss on the dust surface. In order to compensate the electron loss by stronger ionization, and thus to sustain the plasma, the electron energy increases up to 2.4\,eV. At that stage the HMDSO fragments are involved in the dust growth and their contribution to the plasma maintenance is reduced. It is worth noting here that because the main channel of HMDSO fragmentation in the plasma goes through dissociative-ionization processes involving electrons, an increased electron energy would lead to stronger HMDSO fragmentation and result in smaller chemical species, increasing their variety. When dust escapes the plasma gas-phase, due to different forces exerting on it, the energetic need to sustain the plasma decreases and the electron energy goes down to 2.0\,eV. The loss of electrons on the dust surface becomes limited and the electron number density increases up to 1.2$~\times$~10$^{10}$ cm$^{-3}$. Moreover without dust in the plasma, the HMDSO fragments which generally possess ionization potentials much lower than the one of argon largely contribute to the ionization and thus to the increase of the electron number density.
The different chemical environments and the variety of chemical species during different stages of dust formation and escape from the plasma gas-phase determine the dynamical character of the chemical reactions and increase their possible pathways.

The complex chemistry and the large number of elementary processes occurring in the plasma  define the total absorbed power per unit area S$_{abs}$\,(W/m$^{2}$). They are also at the origin of the differences in S$_{abs}$ although the injected power is kept constant (30\,W) for the series of performed experiments. The total absorbed power in this kind of plasma is a sum of two components, each one representing the Ohmic and the stochastic electron heating. They both depend on the RF voltage amplitude and are weighed by the energy transfer from the electrons to the atoms and molecules in the plasma. In asymmetric capacitively-coupled plasmas, the self-bias voltage (in modulus) is close to the RF voltage amplitude; being smaller by a factor of 0.8 \citep{lieberman_principles_2005}. The systematic measurement of the self-bias voltage on the powered electrode allows calculation of the amount of total absorbed power per unit area in the plasma. The obtained values are reported in Table~\ref{tab:samples}. The differences in  S$_{abs}$ for the different samples point to variations in the plasma response in relation with its energetic conditions (electron energy and density variations).

\begin{figure}[h!]

\includegraphics[width=0.4\columnwidth]{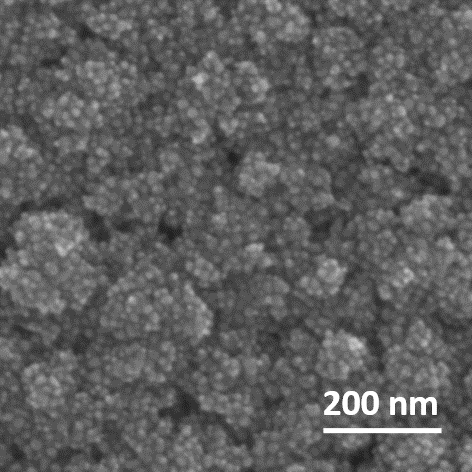}

\caption{Scanning electron microscopy (SEM) image: Plan view of the dust structure corresponding to sample P2, showing the two phases, organosilicon dust and silver nanoparticles organized in a “raspberry-like” structure.}\label{fig:MEB}
\end{figure}

The dust obtained in the plasma gas-phase under the above described dynamic chemical environment is composed of organosilicon (size 200\,nm or less) and  silver (polycrystalline nanoparticles of size 15\,nm) parts. A clear segregation of these two phases is observed. The small whitish dots observed in Figure~\ref{fig:MEB} are identified as silver compounds and the grey surrounding matter is of organosilicon composition as determined by energy dispersive X-ray spectroscopy.
The morphology of the generated dust is of "raspberry-like" type as illustrated on Figure~\ref{fig:MEB},  which suggests a highly porous structure of the dust particles and cavities in between them.  Further results on these aspects can be found elsewhere \citep{Berard21}. For comparison reasons, an organosilicon only dust sample {\it i.e.}, without the silver component (called hereafter reference sample), is included in the discussion of the results. The reference sample was obtained after injection in the plasma of a larger amount of HMDSO.  The exact plasma operating conditions for generation of this sample are given in \cite{Berard19}. Table~\ref{tab:samples} presents a summary of the plasma samples used in this work.\\

\subsection{Dust infrared spectroscopy}\label{Sec:IRspectra}

\begin{figure}
\centering
\begin{tabular}{cc}
\multicolumn{2}{c}{\includegraphics[height=7cm]{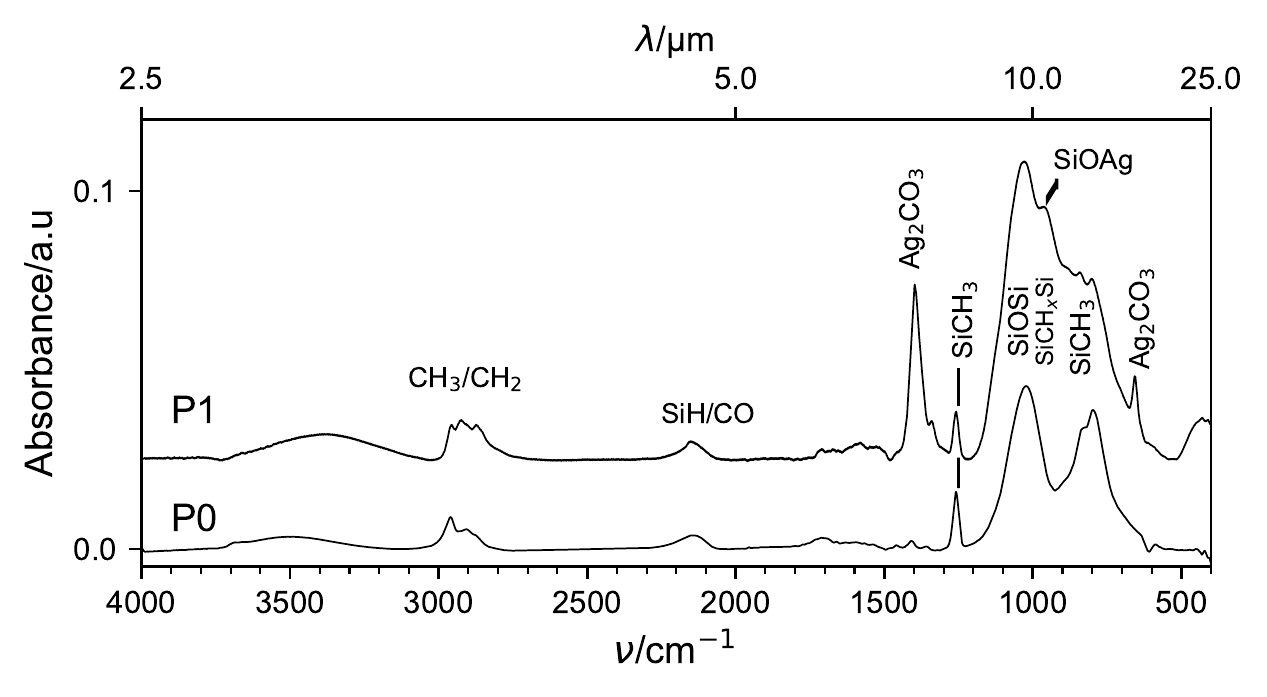}}\\
\multicolumn{2}{c}{\bf{(A)}}\\
\includegraphics[height=6.5cm]{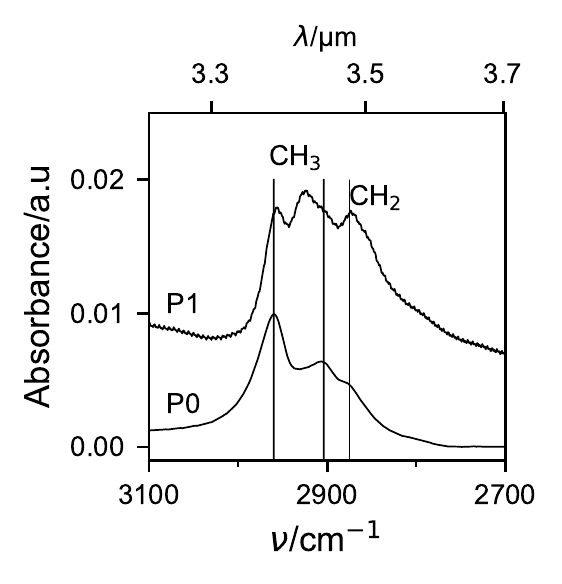}&
\includegraphics[height=6.5cm]{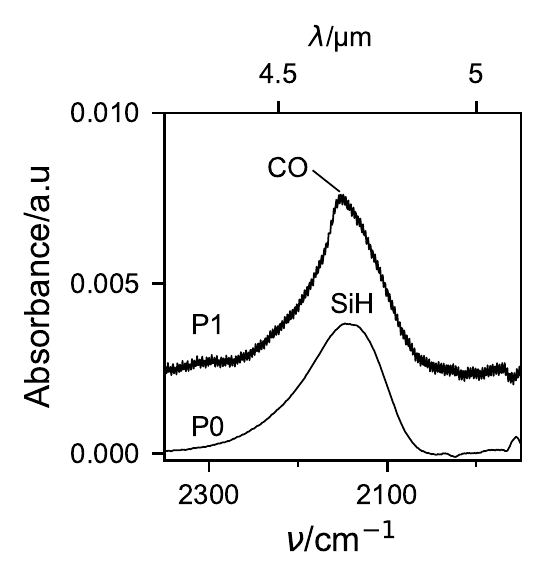}\\
\bf{(B)}&\bf{(C)}\\
\includegraphics[height=6.5cm]{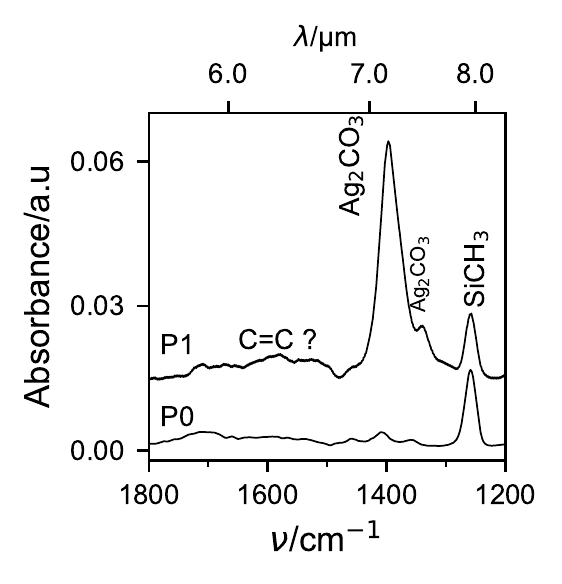}&
\includegraphics[height=6.5cm]{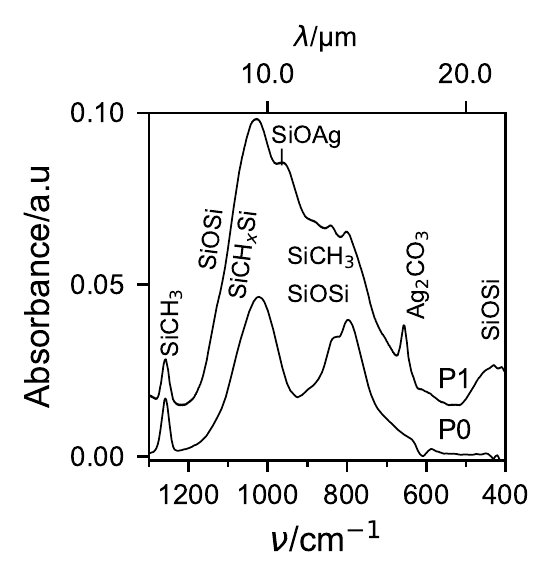}\\
\bf{(D)}&\bf{(E)}\\
\end{tabular}

\caption{ESPOIRS-FTIR spectra of the collected dust in plasma experiments. {\bf(A)} Full IR spectrum for samples P0 (without silver) and P1 (with silver) with the principal band attributions indicated (see text for more details). Zooms in on the most interesting spectral regions are presented in panels {\bf(B)}, {\bf(C)}, {\bf(D)} and {\bf(E)}}\label{FigIR}
\end{figure}

\begin{figure}
\centering
\begin{tabular}{cc}
{\includegraphics[height=6.5cm]{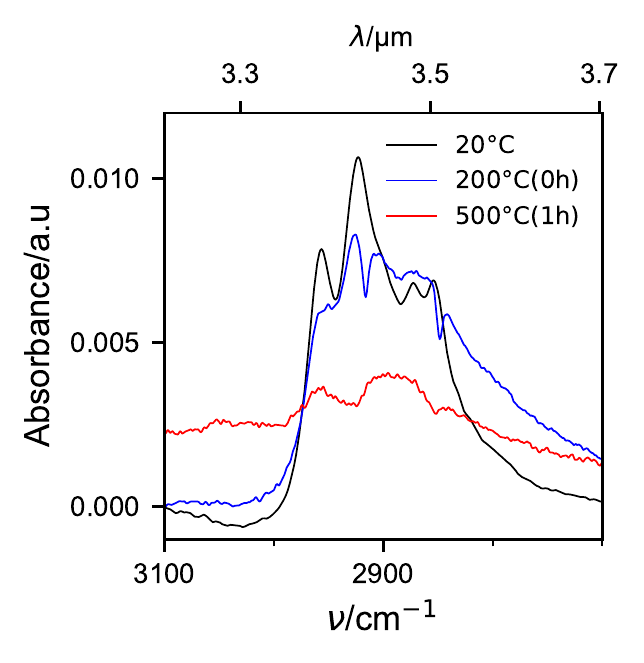}} &
{\includegraphics[height=6.5cm]{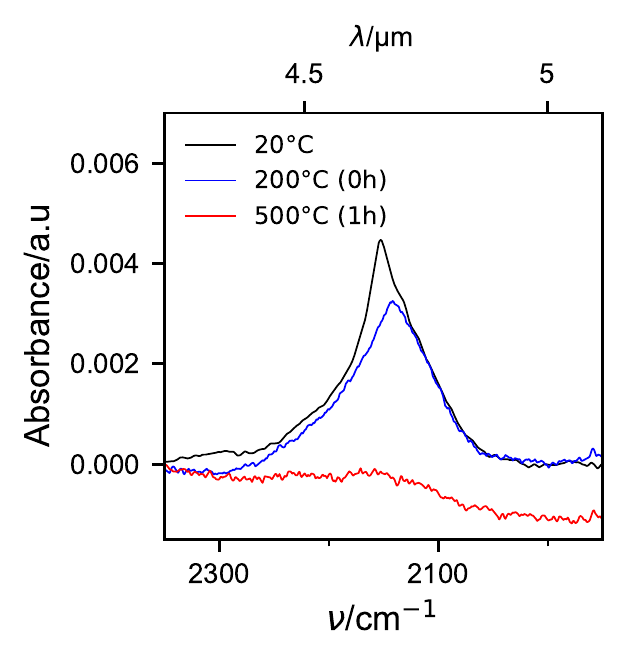}}\\
{\bf{(A)}}&{\bf{(B)}}\\
&\\
{\includegraphics[height=6.5cm]{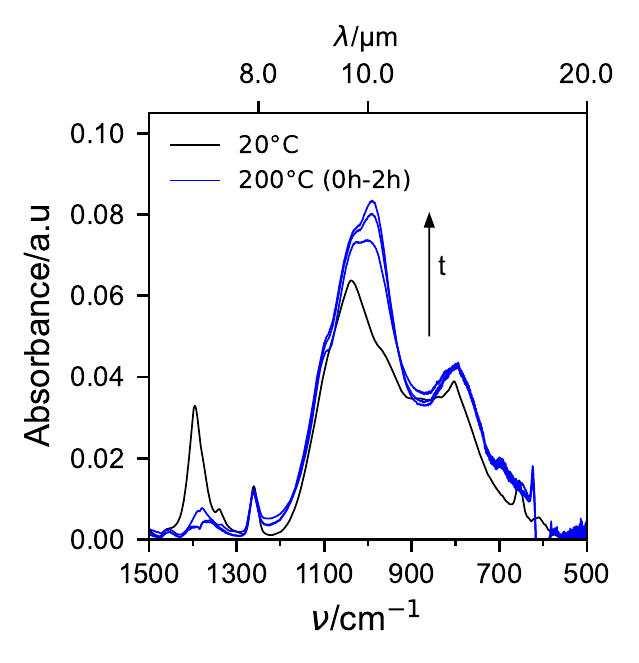}}&
{\includegraphics[height=6.5cm]{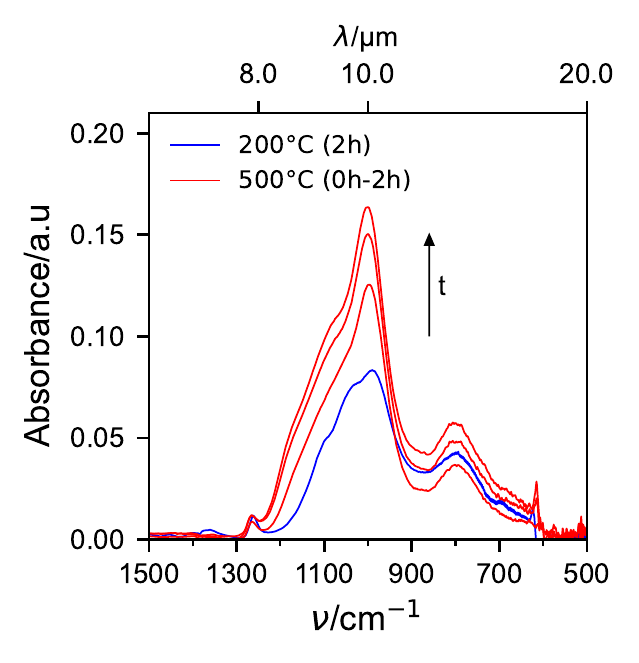}}\\
{\bf{(C)}}&{\bf{(D)}}\\
\end{tabular}
\caption {Evolution of the most interesting bands presented in Figure~\ref{FigIR} with time at 200$^{\circ}$C and 500$^{\circ}$C. For each temperature the presented spectra are recorded at t=0, 1, 2\,h. The spectrum at 20$^{\circ}$C is also plotted for comparison.}\label{FigIR2}

\end{figure}

Infrared spectra are presented in Figure~\ref{FigIR} for the plasma reference sample (P0) produced without silver and for the sample (P1) produced with silver in the plasma.
The IR spectrum of P0 shows the characteristic bands of an organosilicon structure (Figure~\ref{FigIR}~(A)). 
One can observe the two typical broad massifs centered around 800\,cm$^{-1}$ and 1000\,cm$^{-1}$ (Figure~\ref{FigIR}~(E)) that indicate the amorphous structure of the organosilicon dust through the large band width. The intense band at 1030\,cm$^{-1}$ can be assigned to Si-O-Si asymmetric stretching mode in a silicon suboxide environment with a contribution of the Si-CH$_{x(x \le 2)}$-Si wagging mode, most probably positioned at around 1020\,cm$^{-1}$. The band in the 800\,cm$^{-1}$ range is likely a mix of the Si-O-Si symmetric stretching vibration at 810\,cm$^{-1}$ with contribution of the symmetric stretching vibration of Si-C usually appearing at 840\,cm$^{-1}$, as well as several other bands between 700 and 900\,cm$^{-1}$ associated with Si-C and Si-O bonds of the organosilicon structure \citep{smith_infrared_1960,benitez_improvement_2000,raynaud_ftir_2005,despax_deposition_2007, li_thermal_2020}.
Amongst the minor features, methyl CH stretching modes can also be observed at 2960\,cm$^{-1}$ and 2904\,cm$^{-1}$, together with the typical symmetric bending  mode of Si-CH$_{3}$ at 1260\,cm$^{-1}$, which is related to the decomposition of HMDSO in the plasma. The band at 2876\,cm$^{-1}$ due to the symmetric stretching of CH$_2$ is also present on the spectrum. It is paired with the Si-CH$_2$-Si bridge (bending mode) at 1358\,cm$^{-1}$, which is very slightly visible in the spectrum.

With the addition of silver in the experiments (sample P1), new bands appear in the spectrum. It is worth noting that the spectra of all silver-containing dust samples contain identical information. They all exhibit the same spectral bands although their relative intensities may vary slightly from one sample to another reflecting the heterogeneity of the produced dust. The most intense band is observed at 1395\,cm$^{-1}$ with a shoulder at 1340\,cm$^{-1}$. Another clear new band is observed at 655\,cm$^{-1}$. The 1395\,cm$^{-1}$ band has been attributed to the asymmetric stretching of CO$_{3}$ in silver carbonate (Ag$_2$CO$_3$) and the band at 655\,cm$^{-1}$ to its rocking mode \citep{gatehouse_carbonate_1958,slager_infrared_1972}. We also observed a band at 883\,cm$^{-1}$, heavily blended with other nearby OSiO bands, which could correspond to the CO$_3$ out-of-plane deformation mode \citep{slager_infrared_1972}. To get further confidence into this assignment, we have measured with ESPOIRS the spectrum of bulk silver carbonate. It shows a good correspondence within $\sim$30\,cm$^{-1}$ for the main band at 1395\,cm$^{-1}$ and its shoulder \citep[see][]{berard2019phd}. The band at 655\,cm$^{-1}$ is less concordant with the spectrum of bulk Ag$_2$CO$_3$ but the agreement improves when compared to the spectrum of Ag$_2$CO$_3$ thin film \citep{slager_infrared_1972}, with the latter being more relevant to our study. 

The band at 2144\,cm$^{-1}$ (Figure~\ref{FigIR}~(C)) observed in the Ag-free sample (P0) spectrum and attributed to Si-H stretching mode is modified in the presence of Ag. In the spectrum of the Ag-containing sample (P1), the band shape has changed with a maximum of the band at 2153~\,cm$^{-1}$. The position and shape of the SiH band are found to depend on the chemical environment of the grain \citep{Moore91} and also on the gas-phase environment through gas-grain interactions \citep{Accolla21}. The observed variations could therefore be attributed to this effect. However we found that another scenario might be more attractive. Indeed, experiments on CO adsorption on water ices show the presence of a specific band at 2152-2153~\,cm$^{-1}$ in the case of CO adsorption in pores compared to a band at 2139~\,cm$^{-1}$ for surface adsorption sites \citep{Fraser04, Taj17, Taj19}. The relatively sharp band that we observed at 2153~\,cm$^{-1}$ might well be due to CO entrapped in the porous dust (cf. Section~\ref{Sec:Dust}).

Another new band, which cannot be attributed to carbonate nor to adsorbed CO is observed at 965\,cm$^{-1}$ (Figure~\ref{FigIR}~(E)), blended with the bands of the asymmetric stretching mode of Si-O-Si band and the Si-CH$_{x(x \le 2)}$-Si wagging mode at around 1020\,cm$^{-1}$. The assignment of this band is debated in the literature. It has been reported in several studies and attributed either to Si-OH or Si-O-Ag bond. A band at this position is observed in metal-free samples in which it is attributed to SiOH \citep{racles_silica-silver_2013}. On the other hand the infrared spectrum of amorphous AgSiO measured by \cite{cao17} shows a broad band at 987~\,cm$^{-1}$, also observed in Ag$_6$Si$_2$O$_7$ \citep{Qin19}, close to the band observed in the spectrum of sample P1.
The band at $\sim$ 450\,cm$^{-1}$ present in the sample containing silver is due to Si-O-Si rocking vibrations \citep{kirk_quantitative_1988, innocenzi_orderdisorder_2003}. This band cannot be specifically attributed to Ag-containing samples  since it is in general also present in the organosilicon phase, although weak \citep{Mak16}. This band is not seen here in the P0 sample, which can be explained by its intensity depending on the plasma conditions and on the oxygen content \citep{Mak16, berard2019phd}.

Comparison of the two spectra shows that the massif corresponding to the methyl stretching modes around 2960\,cm$^{-1}$ is modified by the addition of silver in the experiment. Some new bands appear at shorter wavenumbers in the spectrum of sample P1 and the band ratios change (Figure~\ref{FigIR}~(B)). The band at 2925\,cm$^{-1}$ (symmetric stretching) becomes more intense than the band at 2960\,cm$^{-1}$ (asymmetric stretching) in sample P1. At the same time the CH$_2$ band at 2876\,cm$^{-1}$ is much stronger relative to the others in sample P1. These changes could point to the presence of additional carbonaceous components in the silver-containing sample. The spectral region 1500 - 1800\,cm$^{-1}$ shows many weak and blended bands in both samples (Figure~\ref{FigIR}~(D)). However, two weak bands at 1530 and 1586\,cm$^{-1}$ are clearly observed in sample P1 but not in sample P0, or much weaker. Although it is difficult to identify confidently these bands, their presence in sample P1 is compatible with an additional carbonaceous component.\\

The evolution with temperature of the spectrum of silver-containing dust is presented in Figure~\ref{FigIR2} in the 500-1500\,cm$^{-1}$ range. Below 600\,cm$^{-1}$, the transmission of the ZnSe windows of the environmental cell was found to be degraded and data at frequency lower than 500\,cm$^{-1}$ are therefore not shown. At 200$^{\circ}$C, the main band of Ag$_2$CO$_{3}$ has disappeared which is consistent with the thermal decomposition of Ag$_2$CO$_{3}$ \citep{ koga_thermal_2013}. Similarly, the sharper component at 2153~\,cm$^{-1}$ (Figure~\ref{FigIR2}~(B)) is not seen at 200$^{\circ}$C, which supports the idea that this band is associated to CO entrapped inside the cavities present in the dust sample. The broad massif around 1000\,cm$^{-1}$ strongly evolves with the temperature increase (Figure~\ref{FigIR2}~(C) and (D)). The contribution to this band of the Si-C-Si stretching vibration at 1020\,cm$^{-1}$ becomes dominant compared to the Si-O-Si asymmetric stretching vibration at 1030\,cm$^{-1}$ for 200$^{\circ}$C, thus placing the peak of this massif at lower energy. After two hours at 200$^{\circ}$C, the massif exhibits three components and its integrated intensity has increased indicating that the Si-O-Si network rearranges itself. This evolution continues with the temperature increase up to 500$^{\circ}$C at which the spectrum exhibits features closer to those of silica but with strong mixing between the Si-O-Si and Si-C-Si vibrations, resulting in a peak at 1000\,cm$^{-1}$ and a well defined band due to the symmetric Si-O-Si stretching vibration at 810\,cm$^{-1}$ \citep{kirk_quantitative_1988, innocenzi_orderdisorder_2003,li_thermal_2020}. Besides, the Si-O-Ag shoulder at 965\,cm$^{-1}$ becomes an integral contribution to the massif. This spectral evolution is most likely due to a further rearrangement of the dust network and to an intensive hydrogen out-diffusion, with the latter expected to start at temperatures higher than 350$^{\circ}$C \citep{wang_effect_1987}. The loss of the SiH band at 2144\,cm$^{-1}$ while heating to 500$^{\circ}$C (Figure~\ref{FigIR2}~(B)) supports the strong decrease in hydrogen concentration, which is expected to lead to the creation of new bonds in the network, such as Si-O, Si-C, Ag-Si, Ag-C, C=C and Si-Si, in order to satisfy the valence bonding of each atom. The creation of new bonds together with the rearrangement of the dust network can account for the increase of the intensity of the spectral massif at $\sim$ 1000\,cm$^{-1}$ from 200$^{\circ}$C to 500$^{\circ}$C. Indications of remaining carbon bonds in the sample at 500$^{\circ}$C are given by the still broad massif peaking at 1000\,cm$^{-1}$ and the band of SiCH$_{3}$ at 1260\,cm$^{-1}$. The remaining signal at $\sim$2950\,cm$^{-1}$ (Figure~\ref{FigIR2}~(A)) is likely to be dominated by the latter bonds. Other CH aliphatic components associated with volatile hydrocarbons have disappeared during the thermal processing.
\\

\begin{figure}[ht!]
\centering
\begin{tabular}{c}

{\includegraphics[width=0.9\columnwidth, height=6.5cm]{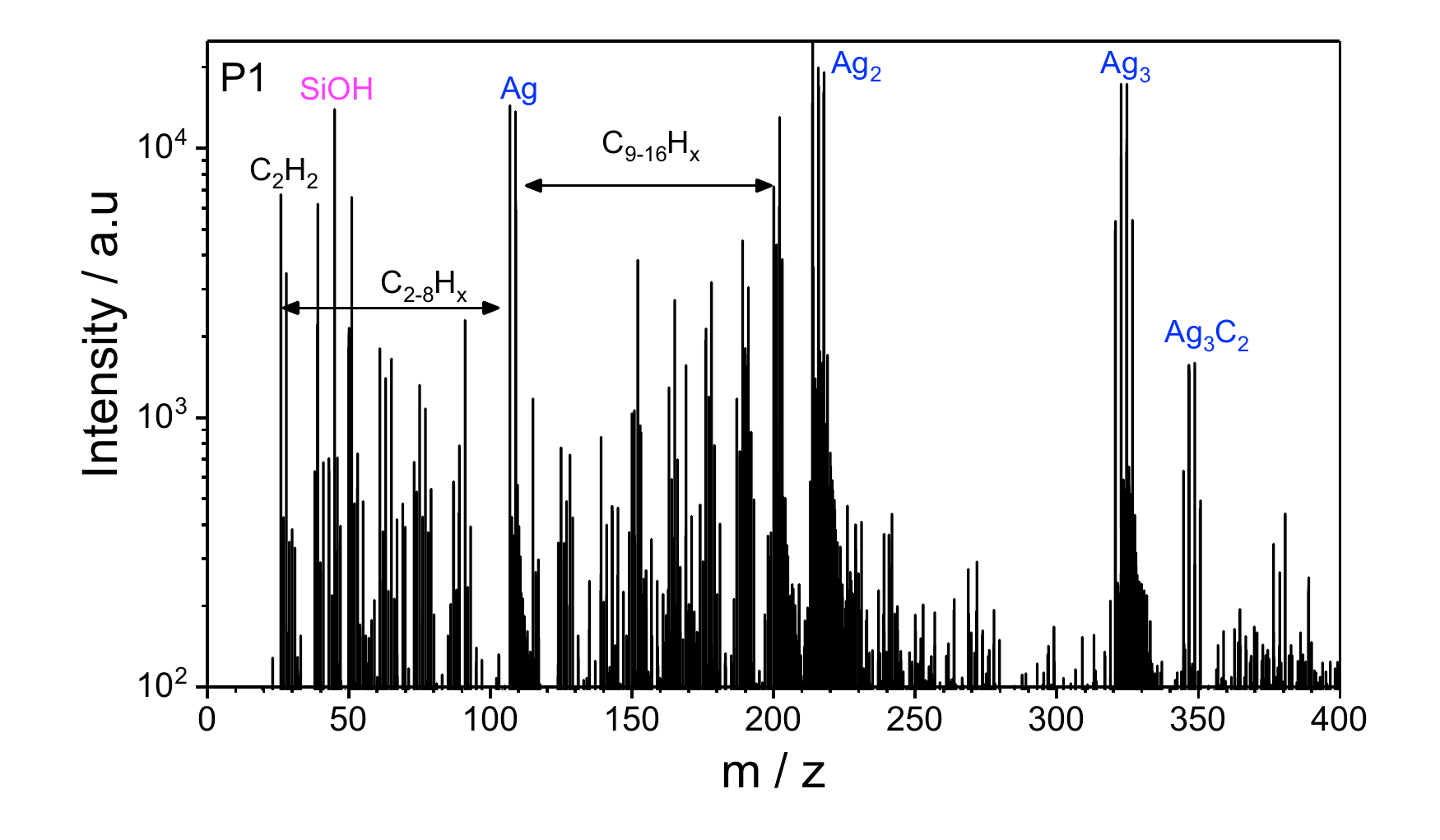}}\\
{\bf (A)} \\

{\includegraphics[width=0.9\columnwidth, height=6.5cm]{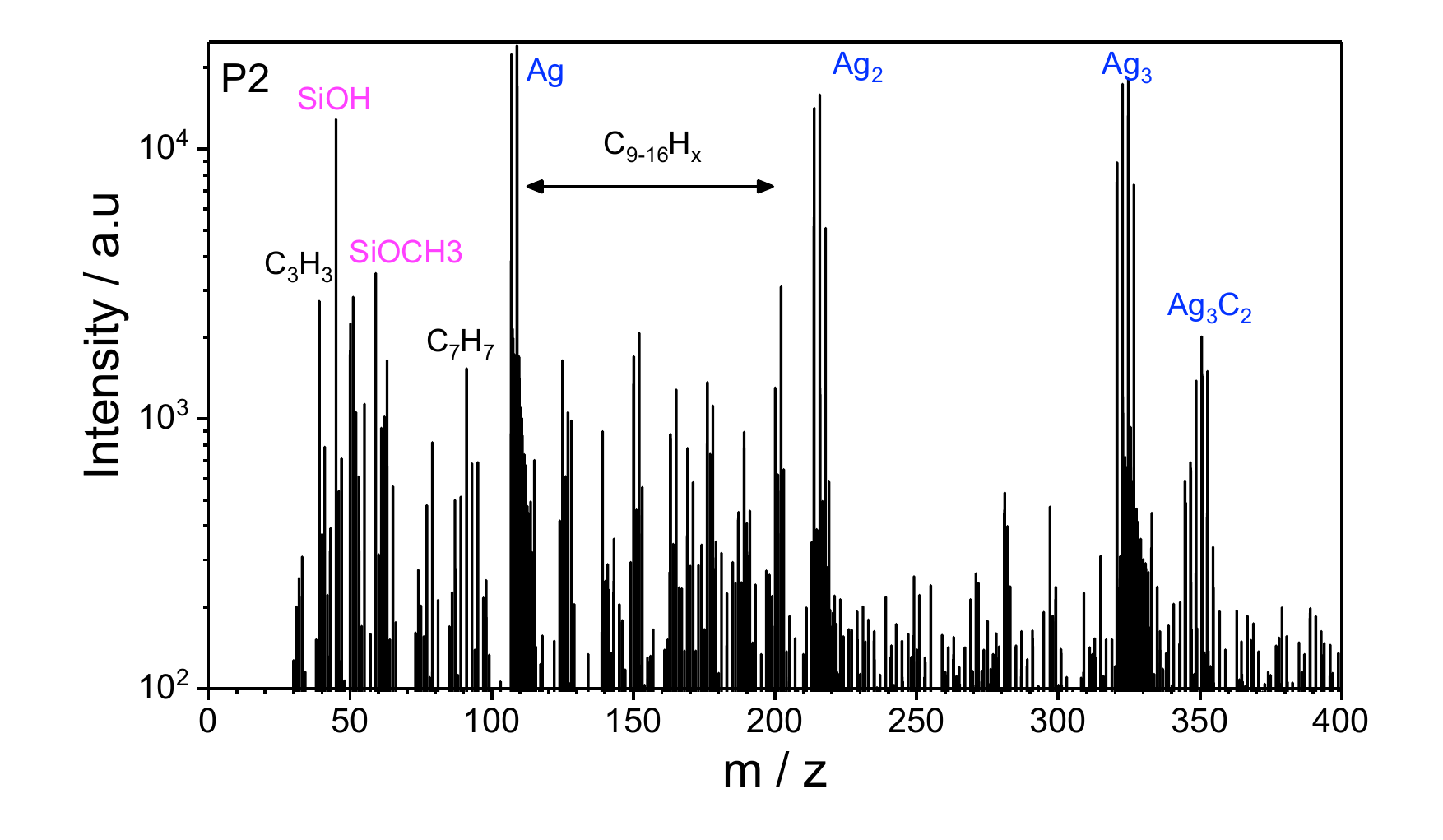}}\\
{\bf (B)} \\

{\includegraphics[width=0.9\columnwidth, height=6.5cm]{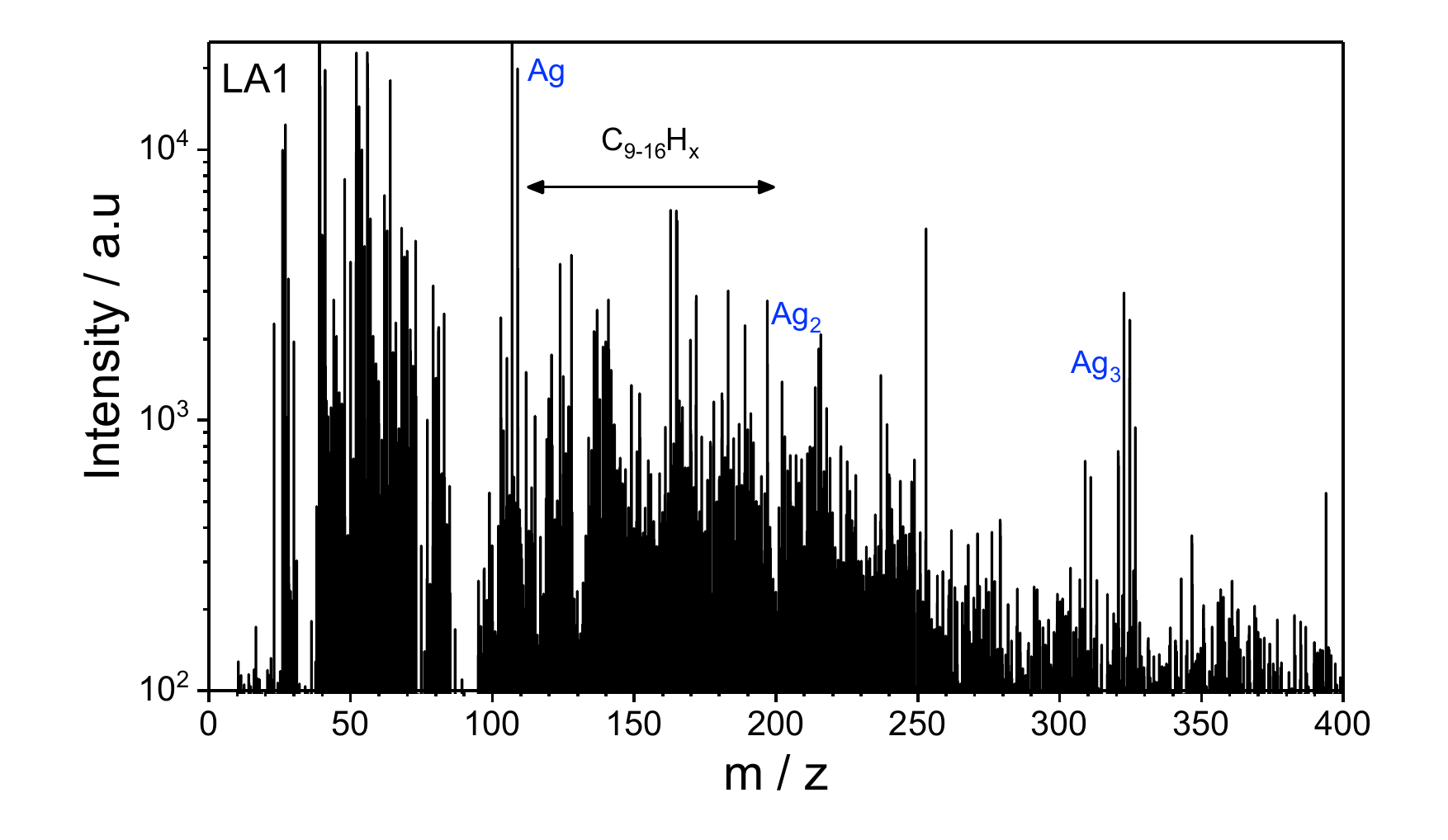}}\\
{\bf (C)} \\
\end{tabular}
\caption{Mass spectra obtained with the AROMA setup for samples P1 (A), P2 (B) and LA1 (C) compiled in Table~\ref{tab:samples}.}\label{fig:mass}
\end{figure}

\begin{table}
\begin{tabular}{|c|c|c|c|c|c|c|}
\cline{2-7}
\multicolumn{1}{c|}{}&\multirow{6}{*}{Silver}&\multicolumn{3}{c|}{\multirow{3}{*}{Silver/Hydrocarbon complexes}}&\multirow{6}{*}{Main hydrocarbons}&\multirow{6}{*}{Silicon}\\
\multicolumn{1}{c|}{}&&\multicolumn{3}{c|}{}&&\\
\multicolumn{1}{c|}{}&&\multicolumn{3}{c|}{}&&\\
\cline{3-5}
\multicolumn{1}{c|}{}&&\multirow{3}{*}{C\#$=$2}&\multirow{3}{*}{H\#$<$(C\#$>$2)}&\multirow{3}{*}{H\#$>$(C\#$>$2)}& &\\
\multicolumn{1}{c|}{}&&&&& \multicolumn{1}{c|}{for a given C\#}&\\
\multicolumn{1}{c|}{}&&&&&&\\
\hline
\multirow{12}{*}{Plasma}&&&&&&\\

&Ag&AgC$_2$H$_2$ &AgC$_3$H$_2$ & AgOC$_3$H$_6$ &C$_2$H$_{2}$,C$_3$H$_{3}$&Si\\
&Ag$_2$& Ag$_2$C$_2$H &AgC$_4$H$_2$&AgC$_5$H$_6$ &C$_4$H$_{3}$,C$_5$H$_{3/5}$&SiCH$_3$\\

&Ag$_3$&Ag$_2$C$_2$H$_2$ &AgC$_4$O&Ag$_2$C$_7$H$_{15}$&C$_6$H$_{5}$ (C$_6$H$_{7}$O), C$_7$H$_{7}$&SiOH\\
&Ag$_4$ & Ag$_2$C$_2$H$_3$  & AgC$_6$H$_4$ & Ag$_2$C$_8$H$_{21}$ & C$_9$H$_{7}$ (C$_9$H$_{6}$O)&SiOCH$_{5}$\\

&Ag$_5$& Ag$_3$C$_2$ & AgC$_7$H$_6$ & Ag$_2$C$_9$H$_{21}$ & C$_{10}$H$_{8/6}$ (C$_{10}$H$_{9}$O)&SiC$_5$H$_{5}$\\
&&  & AgC$_{12}$H$_4$ & & C$_{11}$H$_{7}$, C$_{12}$H$_{8}$ (C$_{12}$H$_{9}$O)&SiC$_9$H$_{7}$\\

& & & AgC$_{13}$H$_6$ & & C$_{13}$H$_{9}$ (C$_{13}$H$_{9}$O)&\\
& &  & AgC$_{13}$H$_8$ & & C$_{14}$H$_{8/10}$ (C$_{14}$H$_{9}$O)& AgSiO\\

& &  & AgC$_{13}$H$_{12}$ &  & C$_{15}$H$_{9}$, C$_{16}$H$_{10}$ (C$_{16}$H$_{11}$O)&Ag$_3$Si\\
&& & AgC$_{14}$H$_{2}$& & & Ag$_3$Si$_2$\\
&&&&&&Ag$_5$Si\\
&&&&&&\\
\hline
\multirow{8}{*}{LVAP}&&&&&&\\

&Ag & Ag$_2$C$_2$H$_2$ & & &C$_2$H$_{2}$, C$_5$H$_{9}$&\\
&Ag$_2$ & Ag$_2$C$_2$H$_4$ &&&C$_6$H$_{7/9}$, C$_8$H$_{9}$&\\

&Ag$_3$ & Ag$_3$C$_2$&&&C$_9$H$_{7/11}$, C$_{11}$H$_{9}$&\\

&Ag$_5$ &&&& C$_{13}$H$_{9}$ (C$_{13}$H$_{9}$O, C$_{13}$H$_{11}$O)&\\

&&&&&C$_{14}$H$_{10}$, C$_{15}$H$_{9}$&\\

&&&&&C$_{16}$H$_{10}$&\\
&&&&&&\\
\hline
\end{tabular}
\caption{Census of the main ions (charge is omitted for clarity) detected in the molecular analysis of the plasma and LVAP samples (Table~\ref{tab:samples}) using AROMA. The chemical categories include silver clusters, hydrocarbons (including related oxygenated species), complexes of Ag with hydrocarbons and Si-bearing species.}
\label{tab:AgC}
\end{table}

\begin{figure}[h!]
\includegraphics[width=1.0\columnwidth]{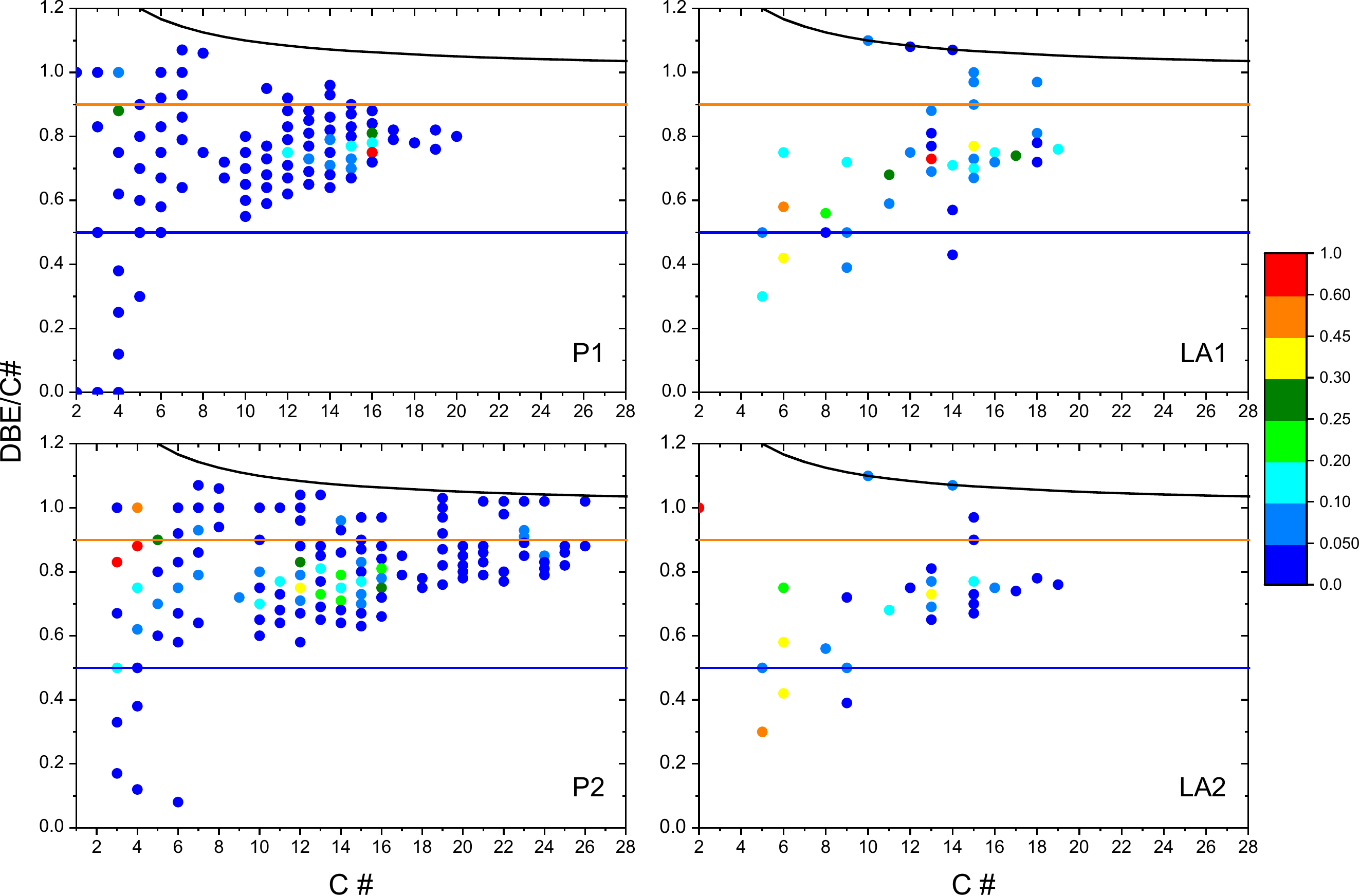}
\caption{Plot of DBE/C\# (see equation~\ref{eq:dbe}) as a function of the carbon number (C\#) for the hydrocarbons and a few carbon clusters, which have been identified with AROMA on samples P1, P2, LA1 and LA2. The corresponding mass spectra for the three first samples are shown in Figure~\ref{fig:mass}. The colored scale refers to the normalized intensity on the most intense peaks in the spectra. Below DBE/C\# = 0.5 the species are considered as aliphatic and above as containing aromatic cycles \citep{Koch2006From}. They remain planar below DBE/C\#=0.9 \citep{Hsu2011}. Above this value, they are classified as HC clusters. The upper black line corresponds to pure carbon clusters.}\label{fig:DBE} 
\end{figure}

\begin{figure}[h!]
\includegraphics[width=1.0\columnwidth]{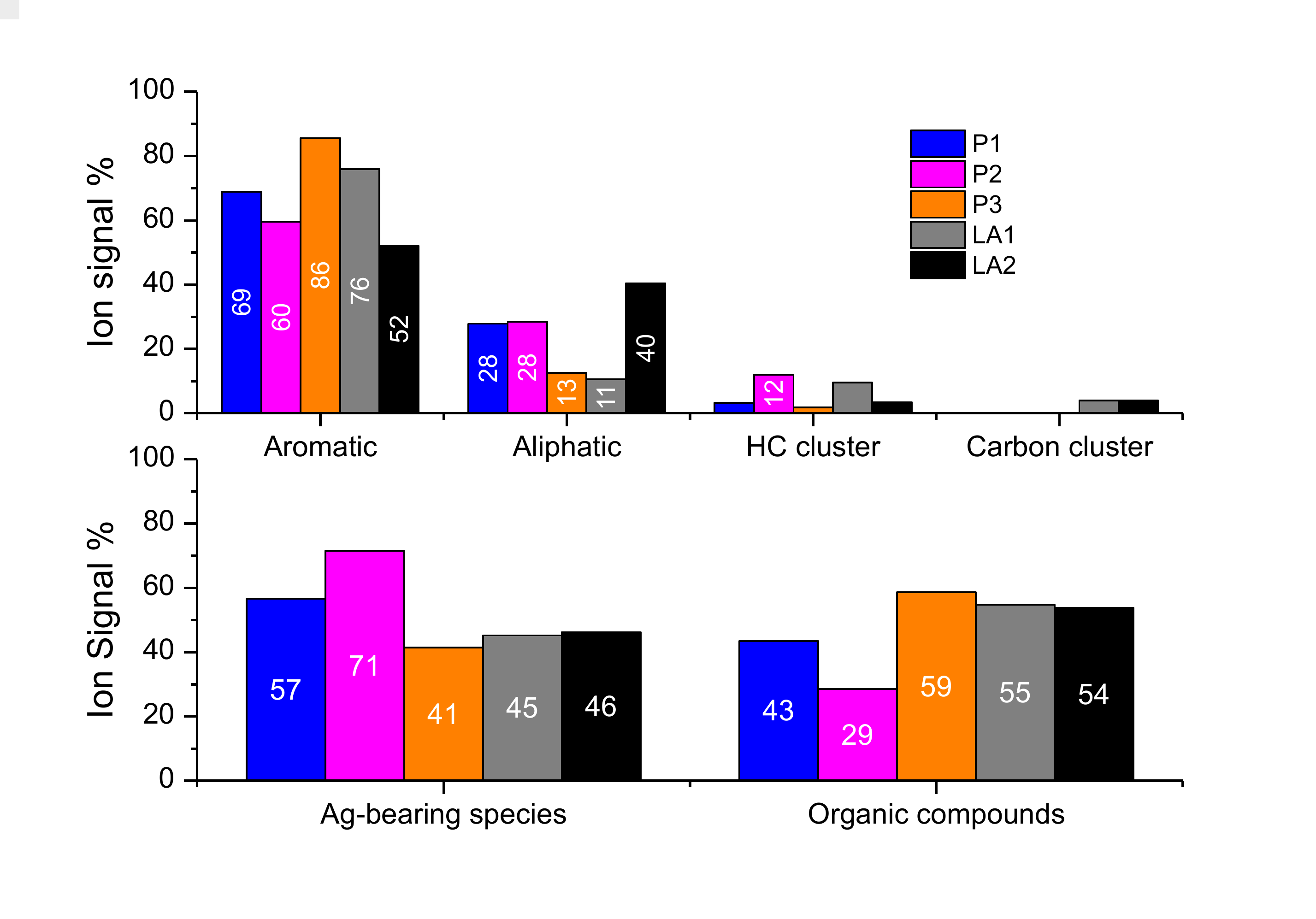}
\caption{Molecular family analysis of the different samples, P1, P2, P3, LA1, LA2 (see Table~\ref{tab:samples}) analyzed with AROMA. The bottom diagram shows the fraction of summed ion intensities of Ag-bearing compounds relative to organic compounds. The top diagram separates the organic compounds (C$_x$H$_y$O$_{0,1}$ species) into the families defined in Figure~\ref{fig:DBE} and Section~\ref{Sec:MolComp}.}\label{fig:pie}
\end{figure}

\subsection{Molecular composition}\label{Sec:MolComp}
Whereas the IR spectra provide global information on the dust composition, the AROMA analysis brings very sensitive information on the molecular content.
Figure~\ref{fig:mass} displays the mass spectra for samples P1, P2 and LA1 (see Table~\ref{tab:samples}), which have been recorded with the AROMA setup in the range m/z=20 to 400. All samples show numerous mass peaks with major peaks corresponding to the cations of Ag, Ag$_2$ and Ag$_3$. In addition, a large variety of peaks are attributed to hydrocarbon species together with a number of mixed silver-carbon species. The attribution of the latter, even if some peaks have low intensity, is favored by the presence of the typical isotope pattern of silver and silver aggregates. The major detected molecular species are listed in Table~\ref{tab:AgC}.
The observed species point to the interaction of C$_2$H$_2$ with silver as discussed in Section~\ref{Sec:Agchem}.
Larger hydrocarbons complexed with silver are also observed and are divided into two categories as a function of their (low versus high) hydrogenation state. Other species observed in the plasma experiments only are also listed in Table~\ref{tab:AgC}. In particular, some peaks are attributed to Si-bearing compounds, such as SiOH or SiOCH$_5$, which are related to the decomposition of the HMDSO precursor. Few silver-silicon mixed species such as Ag$_3$Si and AgSiO can be identified. 

The large amount of different hydrocarbons species is further analyzed. We note that none of these species could be detected for the plasma samples produced without Ag (such as P0), although organosilicon dust was clearly formed. A number of aromatic species are present, up to typically the size of pyrene or its isomers (C$_{16}$H$_{10}$) at m/z=202.08. AROMA is very sensitive to detect these species. Therefore their detection in the plasma samples containing silver, points to a role of this metal in their formation. Aromatic compounds are also clearly detected in the LVAP samples with a number of species in common with the plasma experiments like the strong C$_{13}$H$_9^+$ peak. A noticeable difference though is the non detection of C$_7$H$_7^+$ in LVAP samples, which points to some differences in the chemistry as discussed in Section~\ref{Sec:Cgrowth}. 
 Some oxygenated hydrocarbons are also identified, a large fraction of these corresponding to oxidized aromatic molecules (see Table~\ref{tab:AgC}).
 
In order to further analyze the families of carbonaceous molecules, we used a double bond equivalent (DBE) analysis as described in \cite{Sabbah21}. For a compound of molecular formula (C$_c$H$_h$N$_n$O$_o$S$_s$) the DBE is defined by: \\
\begin{equation}
\rm{DBE = c - h/2 + n/2 +1}
\label{eq:dbe}
\end{equation}

Figure~\ref{fig:DBE} shows the DBE/C\# values as a function of the carbon number (C\#) for all the detected carbonaceous molecules. We do not consider in the analysis the carbon species that are complexed with Ag. The values fall mainly in the range $0.5 \leq DBE/C\# \leq 0.9$, which corresponds to planar aromatic species. The species at $DBE/C\# > 0.9$ are called hydrogenated carbon (HC) clusters. This family can include cumulenes, polyynes or non planar aromatic structures \citep{Hsu2011, Lobodin2012}. The limit at $DBE/C\# = 1$ corresponds to carbon clusters. A few of them were observed in the LVAP experiments (Figure~\ref{fig:DBE}). Finally species with $DBE/C\# < 0.5$ correspond mainly to aliphatic hydrocarbons with $C\#\leq 6$. AROMA is not optimized to detect aliphatic hydrocarbons which have large ionization potentials (see Section~\ref{subsec:method_mole}). Complexation with Ag/Ag$^+$ can however favor their detection \citep{Grace05}. A few species have been evidenced consisting of Ag$_2$ complexes with alkanes (see Table~\ref{tab:AgC} and discussion in Section~\ref{Sec:Cgrowth}).

Figure~\ref{fig:pie} compiles information on the relative contribution in summed ion intensities of Ag-bearing species and hydrocarbons for the different studied samples. One can see that the fraction of Ag-bearing species changes in the different plasma samples from a value of about 70\% to 40\%, whereas it is similar ($\sim$45\%) in the two LVAP samples. On the other hand, the relative summed ion intensities of aromatic relative to aliphatic species changes significantly from a value of $\sim$7 in samples P3 and LA1  to $\sim$2 in samples P1, P2 and $\sim$1 in sample LA2.
One expects a larger amount of Ag in the gas-phase for higher total absorbed power in plasmas or for a higher laser fluence in LVAP experiment. However we could not find a simple relation between the energetic conditions and the amount of Ag-bearing species seen by AROMA. More energetic conditions are also expected to favor aromatic with respect to the more fragile aliphatic species. Although it might be the case in the LVAP experiments, a simple relation was not found in the plasma experiments. All this points to a complex competition between chemical networks.\\

\subsection{Energetics and mechanisms with DFT calculations}
\begin{figure}[h!]
\begin{center}
\includegraphics[width=10cm]{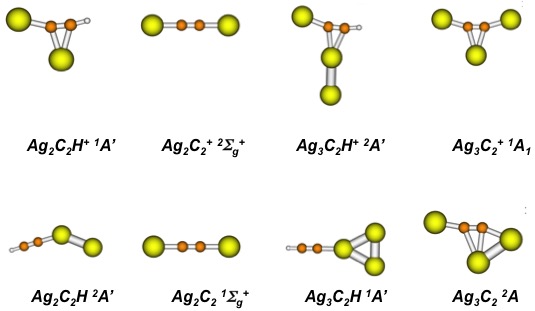}
\end{center}
\caption{Lowest energy structures and electronic states for [Ag$_{2,3}$C$_2$]$^{0/+}$ and [Ag$_{2,3}$C$_2$H]$^{0/+}$ calculated using DFT. }\label{fig:geom_theo}
\end{figure}

In order to rationalize experimental results, structures and energetic data were computed for the smallest observed species {\it i.e.}, Ag$_n$C$_2$H$_m^{0/+}$ (n=1-3, m=0-2). Energetic data include ionization potentials (IPs), adiabatic and vertical, bond dissociation energies (BDEs) and $\Delta H (0K)$ for association reactions. They are reported in Table~\ref{tab:nrj}. The structures of neutral complexes of interest and their cationic counterparts are reported in Figure~\ref{fig:geom_theo}. The determination of these structures has been inspired by DFT calculations previously published in the literature. For Ag$_n$C$_2$H$_m$ neutral clusters, starting point geometries were determined using the results by \cite{DFT_AgnC2H_2013}, who focused on Ag$_n$C$_2$H$^{0/-}$ (n=1-4) complexes, invoking their interest in terms of  catalytic properties. Regarding Ag$_n$C$_2^{0/+}$ complexes, our work was based on the results obtained by \cite{cohen_formation_2011} showing the peculiar stability of M-$\pi$-(M$_2$C$_2$)$^+$ clusters with M=Ag,Au through DFT calculations run in conjunction with high-energy collision experiments.
Interestingly, we found here that the most stable structures for Ag$_n$C$_2$H$^{+}$ (n=2,3) differ from those of their neutral counterparts, leading to important differences between the adiabatic and diabatic (vertical) IPs (8.41 against 5.55\,eV for n=2). Indeed, in the structures of Ag$_n$C$_2$H$^{+}$ (n=2,3), one Ag atom forms a covalent bond with a C atom while the rest of the metallic cluster (Ag or Ag$_2$) has a $\pi$ interaction with the  AgC$_2$H$^{+}$ moiety. This generalizes the statement by \cite{cohen_formation_2011}  to M-$\pi$-(MC$_2$H$^+$) clusters.  

Unless specified, the IP values discussed hereafter are the adiabatic IPs. As can be seen in Table~\ref{tab:nrj}, the IPs of Ag$_n$ species (n=1-3) are found between 5.5 and 7.6\,eV. They are quite similar for n=1,2, and lower for n=3. The values are found in excellent agreement with published experimental values of ionization energies (IEs) for n=1 and 2, and slightly too low for n=3 (see Table~\ref{tab:nrj}).
For n=4 and 5, experimental IEs were measured at 6.65\,eV \citep{jackschath_electron_1992} and 6.35 \citep{jackschath_electron_1992} /5.75\,eV \citep{alameddin_electronic_1992}, respectively.
Several theoretical studies have been dedicated to the geometric and electronic structures of Ag clusters \citep{walsh_s_p_theoretical_1986,mckee_density_2017,sioutis_jahn-teller_2007,tsuneda_theoretical_2019} with the first {\it ab initio} studies on Ag$_2$ and Ag$_3$ in 1986 \citep{walsh_s_p_theoretical_1986}, the latter undergoing Jahn-Teller effect \citep{sioutis_jahn-teller_2007}. Before studying Ag$_n$C$_2$H$_m$ clusters, we checked our level of theory on Ag clusters based on these previous studies. 

The ground states geometric and electronic structures of small neutral and charged organometallic complexes Ag$_x$C$_m$H$_n$ were investigated using the approach detailed in Section~\ref{subsec:method_theo}. The structures and energetics of similar complexes were previously investigated with various DFT functionals and basis sets mostly in synergy with experimental studies \citep{cohen_formation_2011,yang_cluster_2019,tian_experimental_2005,DFT_AgnC2H_2013}. The IPs of Ag$_n$C$_2$H$_m$ (n=1-3, m=0-2) were found to lie between 4.9 and  9.1\,eV. For Ag$_n$C$_2$H$_2$ (n=1-3), the IPs were determined to be lower than those of the corresponding pure Ag$_n$ clusters by -1.32, -0.67 and -0.34\,eV  for n=1-3, respectively. For Ag$_n$C$_2$H, the trend is reverse for n=1,3 as the complexation to C$_2$H leads to an increase of the IP by +1.57 and +1.36\,eV, respectively, with respect to Ag and Ag$_3$. Interestingly, the trend is the reverse regarding the adiabatic IP of Ag$_2$C$_2$H, which was found -2.03\,eV below that of Ag$_2$. The IPs of Ag$_2$C$_2$ and Ag$_3$C$_2$ were found respectively higher (+0.11 eV) and lower (-0.67 eV) than those of the corresponding pure Ag clusters.
The IP of Ag$_3$C$_2$ is the lowest among all computed IPs. It is -2.8\,eV lower than that of Ag$_2$C$_2$, in particular. This is related to the peculiar stability of Ag$_3$C$_2$ \citep{cohen_formation_2011}. 

The metal-ligand bond strengths for [Ag$_n$C$_2$H$_m$]$^{0,+}$ (n=1-3, m=1,2) were  estimated computing adiabatic BDEs. 
Neutral Ag$_n$C$_2$H$_2$ $\pi$ complexes were found to be weakly bound, with BDEs of 0.07, 0.36 and 0.57\,eV for n=1-3, respectively. Larger values of 1.38, 1.04 and 0.91\,eV were obtained for their cationic counterparts for n=1-3, respectively. After losing one H atom, the metal-ligand bond in Ag$_n$-C$_2$H species, which presents a covalent character, is stronger compared to the bond in Ag$_n$C$_2$H$_2$ complexes. BDEs of 3.62, 2.54 and 4.15\,eV were obtained for n=1-3, respectively. The lower value for the even n value can be accounted for by the difference between the open/closed shell characters of Ag$_n$ (closed shell for n=2, open shell for n=1,3) and  Ag$_n$C$_2$H (open shell for n=2, closed shell for n=1,3), the closed shell character leading to increased stability. For [Ag$_n$-C$_2$H]$^+$, BDEs of 2.04, 4.53 and 2.79\,eV were obtained for n=1-3, respectively. 
The higher value found for n=2 can be explained by the closed shell character of [Ag$_n$-C$_2$H]$^+$. The structure of  Ag$_2$C$_2$ differs from that of Ag$_2$C$_2$H as it resembles an acetylene molecule where two hydrogen atoms have been substituted by two Ag atoms, the two Ag atoms are therefore not bounded. The Ag-C BDE within this complex has been estimated to be 3.75\,eV.

\begin{figure}[h!]
\begin{center}
\includegraphics[width=10cm]{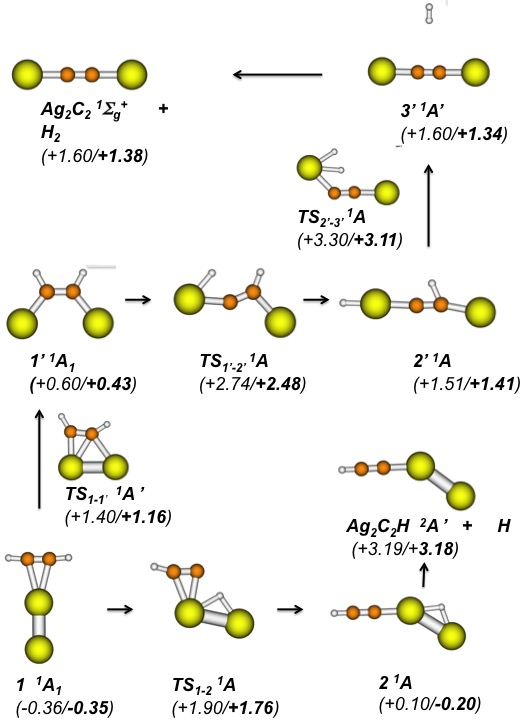}
\end{center}
\caption{Calculated possible intermediates and transition states leading to the loss of H (bottom) and H$_2$ (top)
from Ag$_2$C$_2$H$_2$ 
(structure 1; Ag atoms are yellow). The electronic states are reported, and the DFT relative enthalpies at 0\,K with respect to Ag$_2$ + C$_2$H$_2$ in their ground electronic states are reported in eV in parentheses, along with CCSD(T) single point values in bold characters. 
For TS$_{1-1'}$, no IRC path could be computed due to convergence problem.
}\label{fig:path_theo}
\end{figure}

Regarding the modeling of reactivity, we focused on dehydrogenation reactions starting from the Ag$_2$C$_2$H$_2$ complex, that would lead to a reactive  C$_2$H or C$_2$ site for further carbon growth chemistry (see Section~\ref{Sec:Cgrowth}). As we will see in the discussion section, complexity drastically increases when increasing the size of the system. 
Some possible reactive intermediates and transitions states involved in the dehydrogenation of the Ag$_2$C$_2$H$_2$ complex are reported in Figure~\ref{fig:path_theo}. The loss of H to form the potentially reactive intermediate Ag$_2$C$_2$H is limited by the thermodynamics: it is endothermic by 3.19 (DFT)/3.18 (CCSD(T))\,eV with respect to the ground state of the reactants Ag$_2$ + C$_2$H$_2$, while the H migration from acetylene to Ag$_2$ within the complex only costs 1.90 (DFT)/1.76 (CCSD(T))\,eV with respect to Ag$_2$ + C$_2$H$_2$. 
More favorable pathways involve H abstraction mechanisms. If Ag is involved then 
the intermediate {\bf 2} in Figure~\ref{fig:path_theo} can exchange an H atom with Ag to form Ag$_2$C$_2$H and AgH, this reaction being estimated endothermic by 0.81 eV (see Table~\ref{tab:nrj}), the rate limiting step becoming {\bf 1}$\rightarrow${\bf 2}.
If instead Ag$^+$ is involved, then the formation of the intermediate involves charge transfer and becomes exothermic by 1.15 eV (see Table~\ref{tab:nrj}).

The mechanism for the loss of H$_2$ is expected to be distinct from that of the loss of H, as in the structures of the final products, the atoms have different adjacencies. In the case of Ag$_2$C$_2$H, the identity of the silver dimer is preserved while silver is under the form of two atoms in Ag$_2$C$_2$, with a structure similar to that of acetylene, as specified hereabove. Therefore we considered a mechanism starting from {\bf 1'} in Figure~\ref{fig:path_theo}, with a structure analogous to that of ethylene, located at  0.60 (DFT)/0.43 (CCSD(T))\,eV above the reference corresponding to the separated neutral reactants Ag$_2$ + C$_2$H$_2$.  A transition state {\bf TS$_{1-1'}$} connecting the two minima was computed but its relevance remains uncertain. The path leading to Ag$_2$C$_2$ involves the insertion of an Ag atom into a C-H bond via {\bf TS$_{1'-2'}$} (+2.74 (DFT)/+2.48 (CCSD(T))\,eV above the reference Ag$_2$ + C$_2$H$_2$). Then the second H atom remaining bonded to a C atom migrates onto the Ag atom already carrying the first H atoms to form the {\bf 3'} complex where the H$_2$  molecule presents quasi no interaction with the Ag$_2$C$_2$ moiety. Finally, starting  from {\bf 1'}, the H$_2$ loss reaction is endothermic by $\sim$ 1 eV, the rate limiting step being the last elementary step {\bf 2'}$\rightarrow${\bf 3'}. \\

\begin{table}[ht!]
{\small
\begin{tabular}{|ll|}
\hline
\multicolumn{1}{|l}{Reaction} & \multicolumn{1}{l|}{$\Delta$H(0\,K) (eV)}  \\
\multicolumn{2}{|c|}{Ionization potentials (energies)} \\
Ag ($^2$S) $\rightarrow$ Ag$^+$ ($^1$S) & 7.56 (7.57$^{exp.}$) \\
Ag$_2$ ($^1\Sigma^+_g$) $\rightarrow$ Ag$_2^+$ ($^2\Sigma^+_g$)& 7.58 (7.65) (7.60$^{exp.}$) \\
Ag$_3$ ($^2$B$_2$)$\rightarrow$ Ag$_3^+$ ($^1$A$_1$) & 5.53 (5.82) (5.66$^{exp.\star}$, 6.20$^{exp.}$ ) \\
 AgC$_2$H$_2$ ($^2$A$_1$) $\rightarrow$ [AgC$_2$H$_2$]$^+$ ($^1$A$_1$) & 6.24 (6.36) \\
 AgC$_2$H ($^1\Sigma_g$) $\rightarrow$ [AgC$_2$H]$^+$ ($^2\Sigma_g$) &  9.13 (9.27) \\
 Ag$_2$C$_2$H$_2$ ($^1$A$_1$) $\rightarrow$ [Ag$_2$C$_2$H$_2$]$^+$ ($^2$A$_1$) & 6.91 (7.00) \\
 Ag$_2$C$_2$H ($^2$A')$\rightarrow$ [Ag$_2$C$_2$H]$^+$ ($^1$A') & 5.55 (8.41) \\
 Ag$_2$C$_2$ ($^1\Sigma^+_g$) $\rightarrow$ [Ag$_2$C$_2$]$^+$ ($^2\Sigma^+_g$)& 7.69 (7.83) \\
 Ag$_3$C$_2$H$_2$ ($^2$B$_2$) $\rightarrow$ [Ag$_3$C$_2$H$_2$]$^+$ ($^1$A$_1$) & 5.19 (5.31) \\
 Ag$_3$C$_2$H ($^1$A')$\rightarrow$ [Ag$_3$C$_2$H]$^+$ ($^2$A') & 6.89 (8.33) \\
 Ag$_3$C$_2$ ($^2$A)$\rightarrow$ [Ag$_3$C$_2$]$^+$ ($^1$A$_1$) & 4.86 (4.98) \\
 \multicolumn{2}{|c|}{Association reactions} \\
Ag$_2$C$_2$ + Ag$^+$ $\rightarrow$ [Ag$_3$C$_2$]$^+$ & -3.42 \\
Ag$_2$C$_2^+$ + Ag $\rightarrow$ [Ag$_3$C$_2$]$^+$ & -3.56 \\
 Ag$_2$C$_2$ + Ag $\rightarrow$ Ag$_3$C$_2$ & -0.73 \\
 
   [AgC$_2$H$_2$]$^+$ + Ag $\rightarrow$ [AgC$_2$H]$^+$ + AgH & 2.80 \\
 C$_2$HAg$_2$H (2 in Figure~\ref{fig:path_theo}) + Ag $\rightarrow$ Ag$_2$C$_2$H + AgH & 0.81\\
C$_2$HAg$_2$H (2 in Figure~\ref{fig:path_theo}) + Ag$^+$ $\rightarrow$ [Ag$_2$C$_2$H]$^+$ + AgH & -1.15\\
 
  Ag$_2$C$_2$H + Ag $\rightarrow$ Ag$_3$C$_2$ + H & 2.07 \\
 Ag$_2$C$_2$H$^+$ + Ag $\rightarrow$ [Ag$_3$C$_2$]$^+$ + H & 0.33 \\
Ag$_2$C$_2$H + Ag $\rightarrow$ Ag$_3$C$_2$H &  -2.34\\
  Ag$_2$C$_2$H$^+$ + Ag $\rightarrow$ [Ag$_3$C$_2$H]$^+$ & -1.04 \\
  \multicolumn{2}{|c|}{Bond dissociation energies} \\
AgC$_2$H$_2$ ($^2$A$_1$)$\rightarrow$ Ag + C$_2$H$_2$ ($^1\Sigma^+ _g$) & 0.07 \\

[AgC$_2$H$_2$]$^+$ $\rightarrow$ Ag$^+$ + C$_2$H$_2$ & 1.38 \\
 AgC$_2$H $\rightarrow$ Ag + C$_2$H ($^2\Sigma _g$) & 3.62 \\
 
 [AgC$_2$H]$^+$ $\rightarrow$ Ag$^+$ + C$_2$H & 2.04 \\
  Ag$_2$C$_2$H$_2$ $\rightarrow$ Ag$_2$ + C$_2$H$_2$ & 0.36 \\
  
  [Ag$_2$C$_2$H$_2$]$^+$ $\rightarrow$ Ag$_2^+$ + C$_2$H$_2$ & 1.04 \\
  Ag$_2$C$_2$H  $\rightarrow$ Ag$_2$ + C$_2$H  & 2.54 \\
  
  [Ag$_2$C$_2$H]$^+$ $\rightarrow$ Ag$_2^+$ + C$_2$H & 4.53 \\
  
Ag$_2$C$_2^+$ $\rightarrow$ Ag$_2^+$ + C$_2$ & 5.76 \\
   Ag$_2$C$_2$ $\rightarrow$ Ag$_2$ + C$_2$ & 5.87\\
  
  Ag$_3$C$_2$H$_2$ $\rightarrow$ Ag$_3$ + C$_2$H$_2$ & 0.57 \\
  
  [Ag$_3$C$_2$H$_2$]$^+$ $\rightarrow$ Ag$_3^+$ + C$_2$H$_2$ & 0.91 \\
  Ag$_3$C$_2$H $\rightarrow$ Ag$_3$ + C$_2$H & 4.15 \\
  Ag$_3$C$_2$H $\rightarrow$ Ag$_3$C$_2$ + H  & 4.40 \\
  
 [Ag$_3$C$_2$H]$^+$ $\rightarrow$ Ag$_3^+$ + C$_2$H & 2.79  \\
  
  [Ag$_3$C$_2$H]$^+$ $\rightarrow$ [Ag$_3$C$_2$]$^+$ + H  & 2.37 \\
 
    [Ag$_3$C$_2$H]$^+$ $\rightarrow$ Ag$^+$ + Ag$_2$C$_2$H & 3.00 \\
    
    [Ag$_3$C$_2$H]$^+$ $\rightarrow$ [AgC$_2$H]$^+$+ Ag$_2$ & 3.50 \\
    
    [Ag$_3$C$_2$H]$^+$ $\rightarrow$ Ag$_2^+$ + AgC$_2$H & 1.95 \\
    
Ag$_3$C$_2^+$ $\rightarrow$ Ag$_3^+$ + C$_2$ & 6.54 \\
   Ag$_3$C$_2$ $\rightarrow$ Ag$_3$ + C$_2$ & 5.87 \\
    AgH ($^1\Sigma_g$) $\rightarrow$ Ag + H & 2.28 \\
    C$_2$H$_2$ $\rightarrow$ C$_2$H + H & 5.74 \\
  \hline
\end{tabular}
}

\caption{Thermodynamic data of interest, including BDEs and IPs. The values of the adiabatic IPs are calculated with the structures of the cations which have been optimized, starting from those of the corresponding neutral species.  Vertical IPs are indicated in parentheses. All data are obtained at the DFT level of theory.
The electronic state for each complex is specified once in the table. Experimental ionization energies are retrieved from \cite{jackschath_electron_1992} except for $^{\star}$ which comes from \cite{alameddin_electronic_1992}.
}\label{tab:nrj}
\end{table}

\section{Discussion}

\subsection{Ag$_n$-C$_2$H$_m$ chemistry}
\label{Sec:Agchem}

The precursors used in the experiments is HMDSO in the plasma and acetylene in LVAP, respectively. The HMDSO molecule quickly loses a methyl group after injection in the plasma. Further fragmentation of the precursor in the plasma gas-phase leads to the formation of methane and acetylene, among others \citep{Despax16}. It is therefore not surprising that a number of Ag$_n$C$_2$H$_m^{0/+}$ (n=1-3, m=0-2) species are common for both experiments (see Table~\ref{tab:AgC}).
 DFT calculations contribute to shed light on the specific clusters and molecules that were detected with AROMA. Noticeable species are AgC$_2$H$_2^+$, Ag$_2$C$_2$H$_2^+$, Ag$_2$C$_2$H$^+$ and Ag$_3$C$_2^+$. One should keep in mind that due to the LDI scheme used in AROMA, species with the lowest IPs are expected to be best detected. The much lower IP of Ag$_3$C$_2$ compared to that of Ag$_2$C$_2$ can explain the presence/absence of Ag$_3$C$_2$$^+$/Ag$_2$C$_2$$^+$ in the mass spectra. Ag$_2$C$_2$ might be an important intermediate in the chemistry though if one considers the thermodynamically favorable formation of Ag$_3$C$_2^+$ by association of Ag$^+$ with Ag$_2$C$_2$. Similarly, the adiabatic IP of Ag$_2$C$_2$H is low (5.55 eV) and we observed Ag$_2$C$_2$H$^+$. On the other hand, we could not observe AgC$_2$H$^+$,
 which can be related to the higher IP of its neutral counterpart (9.13\,eV). This scenario is supported by the detection of AgC$_2$H by rotational spectroscopy by \cite{Cabezas2012} after production in a laser ablation source. There is however one noticeable species  for which the argument on the IP fails; Ag$_3$C$_2$H$_2^+$ is not detected in the mass spectra despite Ag$_3$C$_2$H$_2$ has a low IP comparable to that of Ag$_3$C$_2$. 

 This suggests that dehydrogenation has already proceeded from the 
 Ag$_2$C$_2$H$_2^{0/+}$ complexes.  We note that the mechanisms proposed in Figure~\ref{fig:path_theo} were computed for neutral complexes. They are expected to be more energetically favorable for Ag$_2$C$_2$H$_2^+$ complexes as larger metal-ligand  binding energies come into play (see Table~\ref{tab:nrj}). Hydrogen abstraction mechanisms involving Ag atoms/ions are expected to be the most energetically favored pathways. Once formed, the Ag$_2$C$_2$H$^{0/+}$ and  Ag$_2$C$_2^{0/+}$ complexes are expected to be quite resistant even in the energetic conditions of the experiments since their fragmentation involves BDEs of at least 2\,eV. Then larger Ag$_n$ complexes starting from Ag$_3$ can be formed from further Ag$^{0/+}$ addition. We therefore suggest that Ag$_2$C$_2$H$^{0/+}$ and  Ag$_2$C$_2^{0/+}$ complexes are nucleation seeds for Ag$_n$ cluster growth.\\
\\

\subsection{Catalytic role of Ag on hydrocarbon molecular growth in the gas phase}\label{Sec:Cgrowth}

Large hydrocarbons are not present at the beginning of the experiments; in the plasma, methyl groups are abundantly produced from the decomposition of the HMDSO precursor and contribute to form methane and acetylene. Acetylene is injected in LVAP experiments. The AROMA analysis shows that hydrocarbon growth occurs at least up to C{\#}=16 in all experiments (cf. Table~\ref{tab:AgC}). The DBE/C\# diagrams (Figure~\ref{fig:DBE}) show a trend for a periodicity of 2 carbon addition in the LVAP experiments, which is coherent with a growth mechanism based on acetylene. In the plasma, growth also proceeds by single carbon addition, which is expected from the addition of methyl groups. 
There are several reasons why we think that Ag can play a role in hydrocarbon growth. First, the analysis of plasma samples produced in the absence of silver did not provide evidence for the presence of hydrocarbons, in particular aromatic species for which the AROMA analysis is very sensitive \citep{Sabbah2017}. Second, we detected a number of molecular species that are complexes of hydrocarbons with silver (cf. Table~\ref{tab:AgC}).
Third, we observed additional bands in the IR spectrum of the Ag-containing sample compared to the Ag-free sample in the 2800 - 3000~\,cm$^{-1}$ (-CH$_3$, -CH$_2$) and 1500 - 1800~\,cm$^{-1}$ (C=C) spectral regions, which indicates the presence of an additional  volatile hydrocarbon population in the Ag-containing sample (cf. Sect.~\ref{Sec:IRspectra}). In particular, the CH$_2$ band at 2876~\,cm$^{-1}$, which is much more intense in the presence of silver, suggests the presence of long aliphatic species in the Ag-containing sample.
It is important to remind that AROMA provides a biased view of the composition in aliphatic compounds (cf. Sect.~\ref{subsec:method_mole}).
 Although we cannot exclude a contribution of the chemistry on dust surface to the formation of hydrocarbons, we provide below arguments supporting a scenario in which Ag atoms/ions and clusters would promote the growth in the gas phase of hydrocarbons in general and aromatic ones in particular. 
 
 \begin{figure}
\centering
\begin{tabular}{c}

{\includegraphics[height=6cm]{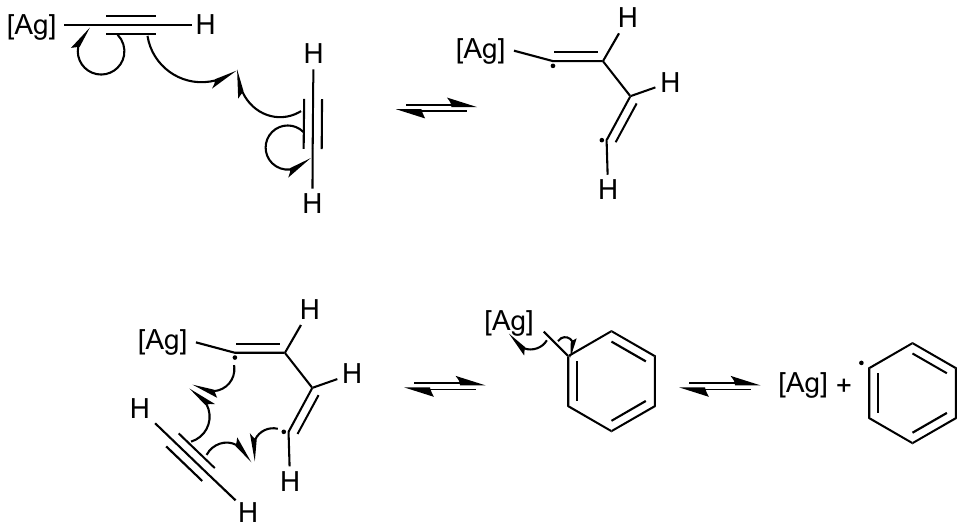}} \\
{\bf (A)} \\
\\

{\includegraphics[height=2.5cm]{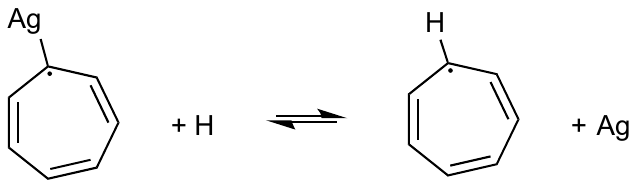}} \\
{\bf (B)} \\
\\

{\includegraphics[height=5cm]{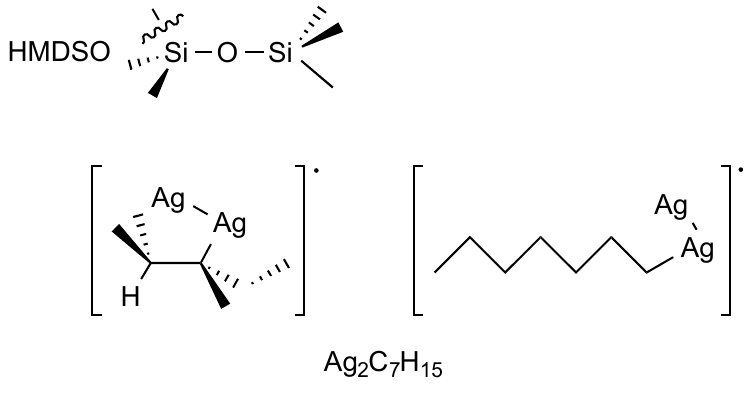}} \\
\\
{\bf (C)} \\
\end{tabular}
\caption{ (A) Possible initial steps of the radical mechanism for C-C coupling on a metallic cluster [Ag] with formation of the first aromatic ring [C$_6$H$_5$]$^\bullet$ ; (B) Substitution of Ag by H to form [C$_7$H$_7$]$^{\bullet}$
(C)~Examples of possible isomers for [Ag$_2$-C$_7$H$_{15}$]$^{\bullet}$. In (A-C) only the H atoms of interest are represented. Let us specify that neutral species are shown here, whereas cationic species are detected by AROMA. }\label{fig:scheme_CCgrowth}
\end{figure}

A bottleneck in the gas-phase formation of aromatics is the formation of the first aromatic ring. \cite{Santoro20} recently carried out experiments on the reactivity of C, C$_2$ and C$_2$H$_2$ in the Stardust machine \citep{Martinez2020}. They concluded that the most likely pathway for benzene formation in these conditions is given by neutral-neutral reactions involving mainly the addition of C$_2$H$_2$ to C$_4$H$_3$ in agreement with reactions derived from dusty plasma \citep{Debleecker06}. Our AROMA analysis reveals indeed the presence of C$_4$H$_3$ and C$_6$H$_5$. In this chemistry, C$_4$H$_3$ needs a termolecular association reaction to be formed:\\
\begin{equation}
\rm{C_2H + C_2H_2 + M  \to C_4H_3 + M}
\end{equation}
with M being the third body, which is necessary to stabilize the reaction complex.
Interestingly, in our case, the reaction could be promoted by complexes of silver with C$_2$H, in particular Ag$_2$C$_2$H, which we have proposed as a seed.\\
\begin{equation}
\rm{Ag_2C_2H + C_2H_2 \to Ag_2C_4H_3}
\end{equation}
Several isomers for  C$_4$H$_3$  are possible, and the most stable cation form differs from the neutral one \citep[see][and references therein]{kaiser_ionization_2010}. For the neutral, a cycle is not the most stable isomer: a linear form (and more precisely  i-C$_4$H$_3$) is the most stable one.   A mechanism we may invoke for carbon growth is a radical mechanism on a metallic center, as illustrated in Figure~\ref{fig:scheme_CCgrowth}~(A),  with Ag$_2$C$_2$H the initial promotor.    
The next C$_2$H$_2$ addition would lead to the formation of Ag$_2$C$_6$H$_5$. The cyclization of C$_6$H$_5$ might be accompanied by a weakening of the Ag$_2$ bonding to the complex facilitating the separation of C$_6$H$_5$ from Ag$_2$ in the gas phase. Further chemistry will then proceed leading to the formation of larger aromatic compounds via C$_2$H$_2$ and CH$_3$ (for the plasma) addition to the first aromatic ring. {The role of metal ions and small clusters in the cyclization of unsaturated hydrocarbons has been previously demonstrated \citep{Schnabel91,Wesendrup97,Chretien03, Marks19}.  We suggest here that Ag$_2$ would act as a catalyst for the formation of the first aromatic ring. 

Another interesting property of silver is the propensity of an Ag atom to substitute to an H atom. This is illustrated in Figure~\ref{fig:scheme_CCgrowth}~(B) for the case of AgC$_7$H$_6$ and C$_7$H$_7$, whose cation is the well known stable tropylium species. The same occurs for the key species C$_4$H$_3$ and C$_6$H$_5$ discussed above, with AgC$_4$H$_2$ and AgC$_6$H$_4$ being identified as main ions in Table~\ref{tab:AgC}. Another noticeable case is AgC$_{13}$H$_8$ / C$_{13}$H$_9$.
The substitution of Ag was also key in revealing the presence of more hydrogenated species, which would form in the plasma experiments but cannot be detected with the AROMA LDI scheme.  For instance, at the bottom of Figure~\ref{fig:scheme_CCgrowth}~(C), two possible stable isomers are reported for  Ag$_2$-C$_7$H$_{15}$. 
In such complexes, it appears unlikely that the carbon structure is cyclic, presenting any aromatic character and Ag$_2$ can be regarded as a substituent of H in heptane C$_7$H$_{16}$. Interestingly, such organometallic complexes are formed on the Ag$_2$ organometallic centers and not Ag. The formation of these species are likely to result from C-C coupling, coupled to H abstraction/H migration, starting from the methyl groups released in the plasma by decomposition of HMDSO.\\

\subsection{Chemical conditions for dust formation}
The formation of dust in the plasma arises from the variety of species produced after fragmentation of the precursor and their further evolution by chemistry. Plasma induced fragmentation of HMDSO provides a very large number of chemical radicals. The main neutral reactive species detected by {\it in situ} mass spectrometry are H, H$_{2}$, C$_{2}$H$_{2}$, CH$_{4}$, CO, CO$_2$ and Si$_{2}$O(CH$_{3}$)$_{5}$, with the latter being characteristic of the injected amount of HMDSO \citep{Despax16}. In addition to the small molecules, one can observe larger mass species, even compounds issued from oligomerization of the parent monomer, and a very rich variety of ions.  The formation of dust in HMDSO-containing plasmas is considered to be based on a negative ion formation scheme, initiated manly by the acetylene-related anions C$_2$H$^-$ and complemented during the dust growth by large C- and Si-bearing species; among others those observed in the plasma are Si(CH$_{3}$)$_{3}$, SiH(CH$_{3}$)$_{2}$, Si$_{2}$OCH$_2$(CH$_{3}$)$_{3}$, Si$_{2}$O(CH$_{3}$)$_{5}$.
In this scheme, the nucleation of dust is limited by the formation and survival of C$_{2}$H$^{-}$, which is formed by electron impact on C$_{2}$H$_{2}$ and reacts quickly with H to reform C$_{2}$H$_{2}$. The dust growth on the other hand depends on the availability of HMDSO fragments. The composition of available fragments depends on the plasma conditions and the quantity of HMDSO injected. For a small amount of injected HMDSO, which is the case for the experiments with silver, the energy of the electrons increases \citep{Berard21}. It is thus expected to lead to a higher fragmentation of the HMDSO {\it i.e.}, to the production of smaller fragments. In particular AROMA evidences the presence of Si and SiO, which shows that the fragmentation can proceed to the atomic and diatomic levels, opening new chemical pathways for dust formation and growth, different than that based on the large fragments of HMDSO. An implication of this chemical complexity is also that the formed dust can be very heterogeneous and with various components located at different scales. \\

\subsection{Ag incorporation in dust}

The main contribution of silver in dust consists in crystalline nanoparticles of diameter $\sim$\,15\,nm \citep{Berard21} that decorate the organosilicon dust particles as shown on the SEM image (Figure~\ref{fig:MEB}). This shows that a large fraction of the Ag-containing dust is segregated from the organosilicon dust. As discussed in Section~\ref{Sec:Agchem}, the formation of silver nanoparticles could be initiated by the formation of Ag$_3$C$_2$H or Ag$_3$C$_2$ complexes that could constitute nucleation seeds for their growth. On the other hand, there is evidence that silver is also involved in bonds with other elements and it is therefore also part of other dust components. {We have also identified in the infrared spectrum a silver carbonate component.}
There is also evidence of silver interaction with SiO, through the $\sim$ 965\,cm$^{-1}$ band, which we assign to  AgSiO.
This result is strengthened by the observation of molecular AgSiO in the AROMA setup. If these species result from dust fragmentation upon laser impact, then it implies, considering the low laser fluences used in AROMA, that AgSiO is part of a highly disordered network. Rearrangement upon annealing and the formation of silica and Ag-silicate structures with increased temperature is therefore expected. At 200$^{\circ}$C, one can note that the massif at $\sim$1000\,cm$^{-1}$ is progressively dominated by the SiOSi band at 990\,cm$^{-1}$.  At 500$^{\circ}$C, this massif narrows showing that the organosilicon dust evolves toward silica (SiO$_2$), even though the structure remains amorphous and with a Si-C rich component (shoulder at $\sim$ 1100\,cm$^{-1}$), both characteristics leading to an increased bandwidth relative to that of crystalline silica. The SiO network rearranges itself and becomes more polymerized. The evolution of the AgSiO phase is more difficult to trace as the 965\,cm$^{-1}$ band is blended with the SiOSi band which increases with temperature from 200 to 500$^{\circ}$C. Although additional experiments would be necessary to conclude on this point, it is likely that the AgSiO component is located at the junction between the silver nanoparticles and the amorphous organosilicon component.\\

\section{Conclusion and astrophysical implications}
The formation of dust in the circumstellar environments of evolved stars is a complicated problem which involves different gas and dust phases in varying physical and chemical conditions \citep{Gail2014book}. There are several aspects that limit the description of dust formation in models. One concerns the possibility to form inclusions in a growing dust particle. In O-rich circumstellar environments, Mg is found to be well incorporated in silicates. This is less clear for iron, which could be mostly included in the form of iron nanoparticles. In their modeling work, \cite{Gail1999} discussed that the formation of iron nanoparticles cannot arise from iron clusters in the gas phase and therefore requires precipitation on grains that are already present. The same reasoning was applied for other phases such as silica.

As a case study, we report here the formation of metallic (silver) nanoparticles (size $\sim$15\,nm) under dusty plasma conditions in which organosilicon dust of size 200\,nm or less is formed. 

These experiments show that the growths of metallic and organosilicon dust parts happen independently, although simultaneously. We found a possible gas-phase route to initiate the formation of Ag clusters, which involves Ag$_2$C$_2$H$^{0/+}$ and  Ag$_2$C$_2^{0/+}$ complexes as nucleation seeds. The growing Ag clusters can create bonds with the organosilicon part leading to junctions, as evidenced by the AgSiO infrared band. Further growth of the Ag clusters can then occur to form the observed polycrystalline silver nanoparticles. For the organosilicon part, trend of organization is observed only upon annealing with formation of a silica-like phase. We could not find clear evidence for the presence of silver silicate, which shows that silver is poorly incorporated in the organosilicon dust. We can conclude that the segregation of iron nanoparticles proposed by \cite{Gail1999} in their model is observed for silver in our experiments, the difference being that the condensation of the silver metallic phase can start in the gas phase. Finally, we also found that, due to its porous nature the formed dust has the ability to adsorb CO, which might be a property to consider in circumstellar environments in which CO is an abundant gas-phase species.

The search for key molecular clusters that would initiate dust formation is at the heart of many studies both from astronomical observations and quantum chemical modeling. Emphasis has been given on the more refractory species that could nucleate the first particles in the expanding circumstellar shells. In the case of O-rich environments, a route to silicates via SiO clusters has been excluded \citep{Bromley2016} and alumina (Al$_2$O$_3$) condensates are considered as possible nucleation seeds \citep{Gobrecht2018}. Models are now developed to include a variety of chemical networks for these clusters \citep{Boulangier2019}. In the case of C-rich environments, the first condensates are expected to be silicon carbide and titanium carbide \citep{Lodders1995}. SiCSi and Si$_2$C have been shown to be the most abundant SiC-containing species in the dust formation zone of the prototype C-rich AGB star, IRC+10216 \citep{Cernicharo2015}. Other seeds could form at lower temperatures while some dust components are already in the growing phase. For instance, we demonstrated the importance of mixed Ag-C molecular species. In particular, we showed that these seeds promoted not only the formation of Ag clusters but also catalyzed hydrocarbon growth.
The role of metals (iron) in gas-phase formation of PAHs} in the envelopes of AGB stars was 
earlier suggested by \cite{Ristorcelli1997} and is seen here in experiments with silver. This organometallic chemistry could be a way to increase the formation rate of PAHs compared to models based solely on hydrocarbon chemistry \citep{Cherchneff92}.

To conclude, the presented experiments and analysis raise the methodology and open perspectives to tackle questions related to dust formation in circumstellar shells of evolved stars. Silver is not a relevant metal in these environments. Still our work provides a proof of concept to study the impact of metallic atoms on the nucleation processes in the ejecta of dying stars. It shows that addressing the problem of dust formation requires a multidisciplinary context. Further experiments are planned to study the case of iron
in an environment in which the key elements involved in stardust formation (C, H, O, Si) are already present.

\section*{Conflict of Interest Statement}

The authors declare that the research was conducted in the absence of any commercial or financial relationships that could be construed as a potential conflict of interest.

\section*{Author Contributions}

RB performed the plasma experiments and the dust characterization. DN-R performed the LVAP experiments. HS, RB, DN-R and FM performed the molecular characterization. RB and KD performed the IR studies. AS made the DFT calculations to elucidate the organometallic chemistry. KM and RB characterized the plasma conditions and CJ and KD the astrophysical context. KM and CJ conceived and planned the experiments, and participated in the data analyses. CJ supervised the project and led the writing of the article. All authors contributed to the manuscript edition and approved the submitted version of the paper.

\section*{Funding}
This work has been funded by the European Research Council under Synergy Grant ERC-2013-SyG, G.A. 610256 (NANOCOSMOS) with PIs Jose Cernicharo, Christine Joblin and Jos\'e \'Angel Mart\'in-Gago.

\section*{Acknowledgments}
This article is dedicated to Dominique Le Quéau, who is deeply acknowledged for his scientific curiosity, enthusiasm and selfless support over many years. All this has been essential in promoting our laboratory astrophysics and interdisciplinary activities.

We acknowledge Jose Cernicharo and Jos\'e \'Angel Mart\'in-Gago for fruitful discussions, as well as
Anthony Bonnamy, Pavol Jusko, Ming Chao Ji and Michel Pellarin for their contribution in the commissioning of the LVAP source. The authors also acknowledge support from UAR Raymond Castaing of the University of Toulouse and Mr St\'ephane Le Blond du Plouy for the SEM observations.

\section*{Supplemental Data}
N/A

\section*{Data Availability Statement}
The mass spectra and associated analysis tools are made publicly available in the AROMA database: http://aroma.irap.omp.eu. The dataset associated with the infrared spectra can be found at http://doi.org/10.5281/zenodo.4544876.

\bibliographystyle{frontiersinSCNS_ENG_HUMS} 

\bibliography{Ag_frontiers}







\end{document}